\documentclass[manuscript,screen]{acmart}

\AtBeginDocument{%
  }



\setcopyright{acmlicensed}
\copyrightyear{2025}
\acmYear{2025}
\acmDOI{XXXXXXX.XXXXXXX}

\acmJournal{TOSEM}
\acmVolume{1}
\acmNumber{1}
\acmArticle{1}
\acmMonth{8}

\usepackage{amsmath,amsfonts}
\usepackage{graphicx}
\usepackage{textcomp}
\usepackage{xcolor}

\usepackage[utf8]{inputenc}
\usepackage{booktabs}
\usepackage{longtable}
\usepackage{xspace}
\usepackage{flushend}
\usepackage{enumitem}
\usepackage[ruled]{algorithm2e}
\usepackage{algpseudocode}
\usepackage{placeins}
\usepackage{color}
\usepackage{makecell}
\usepackage{listings}
\usepackage{bbding}
\usepackage{amsthm}
\usepackage{pifont}
\usepackage{caption} 
\usepackage[normalem]{ulem}
\usepackage[export]{adjustbox}

\usepackage{multirow}
\usepackage{booktabs}
\usepackage{colortbl}
\usepackage{tabularx}
\usepackage[many]{tcolorbox}

\newcolumntype{C}[1]{>{\centering\arraybackslash}p{#1}}
\newcommand{\tool}{\textsc{ProfMalPlus}\xspace}
\newcommand{\todo}[1]{\textcolor{black}{#1}}
\newcommand{\tosem}[1]{\textcolor{black}{#1}}
\newcommand{\modified}[1]{\textcolor{black}{#1}}

\usepackage{nicefrac, xfrac}

 \usepackage{microtype}
\begin{document}

\title{\tool: Agent-Coordinated Detection of Malicious NPM Packages via Static–Dynamic Analysis Synergy}

\author{Yiheng Huang}
\authornote{Y. Huang, Z. Zhao, B. Chen, S. Wu, Z. Zhou, Y. Cao, K. Hu, X. Hu, X. Peng are with the College of Computer Science and Artificial Intelligence.}
\affiliation{
\institution{Fudan University}
\city{Shanghai}
\country{China}
}

\author{Zhijia Zhao}
\authornotemark[1]
\affiliation{
\institution{Fudan University}
\city{Shanghai}
\country{China}
}

\author{Bihuan Chen}
\authornotemark[1]
\authornote{B. Chen is the corresponding author.}
\affiliation{
\institution{Fudan University}
\city{Shanghai}
\country{China}
}

\author{Susheng Wu}
\authornotemark[1]
\affiliation{
\institution{Fudan University}
\city{Shanghai}
\country{China}
}

\author{Zhuotong Zhou}
\authornotemark[1]
\affiliation{
\institution{Fudan University}
\city{Shanghai}
\country{China}
}

\author{Yiheng Cao}
\authornotemark[1]
\affiliation{
\institution{Fudan University}
\city{Shanghai}
\country{China}
}

\author{Kun Hu}
\authornotemark[1]
\affiliation{
\institution{Fudan University}
\city{Shanghai}
\country{China}
}

\author{Xin Hu}
\authornotemark[1]
\affiliation{
\institution{Fudan University}
\city{Shanghai}
\country{China}
}

\author{Xin Peng}
\authornotemark[1]
\affiliation{
\institution{Fudan University}
\city{Shanghai}
\country{China}
}

\renewcommand{\shortauthors}{Huang et al.}

\begin{abstract}
Open source software (OSS) has become the foundation of modern applications, but its transitive dependencies~make~it especially vulnerable to supply chain attacks. One common tactic~is to inject malicious code into third-party packages.~NPM,~in~particular, due to its widespread use and large volume of packages, has become a popular target of malicious code injection.~While various detectors have been proposed, they suffer from \tosem{four} limitations, i.e., inadequate behavior modeling of obfuscated code, ignoring object-centric features of JavaScript, lack of synergy~between static and dynamic analysis\tosem{, and semantic information loss in behavior abstraction as well as a lack of interpretability in detection results}. These limitations lead to imprecise modeling of program behavior and hinder detection effectiveness.
To address these limitations, we propose \tool, \tosem{a malicious NPM package detector that combines behavior graph analysis with coordinated LLM-based reasoning over annotated code slices. \tool first identifies installation commands and entry files, and then constructs object-sensitive behavior graphs to capture sensitive APIs, third-party calls, and unresolved calls. Based on these graphs, \tool extracts security-relevant code slices, and enriches them with inline evidence from static analysis. Given these evidence-enriched slices, local judge agents assess each slice independently, and self-consistency verification consolidates repeated local judgements to reduce LLM variance. A global judge agent then synthesizes the verified local reports into an entry-level verdict. When the entry-level verdict remains undetermined, a router agent selects the next evidence source, i.e., third-party enrichment, which injects registry-derived module and method semantics into the slices, or dynamic augmentation, which executes the package in a sandbox to resolve runtime-dependent or unresolved behaviors. The updated evidence is then fed back into the slices, and the judgement process is repeated. Finally, for confirmed malicious entries, a localization agent reports the concrete malicious code snippets with explanations.}
Our evaluation indicates that \tool achieves the highest F1-score of \todo{98.1\%}, outperforming state-of-the-art detectors by \todo{3.5\%}~to \todo{52.6\%}. In the real-world evaluation, \tool detected \todo{597} previously unknown malicious NPM packages, and all of them have been confirmed and removed from NPM.

\end{abstract}

\maketitle


\section{Introduction}\label{sec:intro}

Open-source software (OSS) constitutes up to 90\%~of~a~modern application’s codebase, with all ecosystem downloads~exceeding 6.6 trillion in 2024~\cite{Sonatype2024report}. Despite this ubiquity,~limited visibility into transitive dependencies and insufficient~investment in software supply chain (SSC) have left OSS vulnerable to certain threats, such as incompatibilities~\cite{jayasuriya2023breakingChanges}, vulnerabilities~\cite{Cao2025Antman}, and malicious packages~\cite{ohm2020backstabber}, increasing SSC risks~\cite{ladisa2023sok}. 

\textbf{Problem.} Malicious packages, which may be delivered~via updates to legitimate libraries~\cite{event-stream} or by publishing new packages that exploit common typos and namespace collisions~\cite{npm_typo}, have emerged as a critical SSC threat. Since November 2023, over 512,847 malicious packages have been identified (a 156\% year-over-year increase)~\cite{Sonatype2024report}, highlighting an escalating wave~of SSC compromise. The NPM registry~\cite{npm_website}, in particular, has seen a marked proliferation of malicious packages, eroding~the security of downstream applications. Given NPM's massive, rapidly expanding registry and ecosystem, manually inspecting newly published packages is infeasible. Therefore, automated malicious package detectors are essential.

\textbf{Limitations of Existing Techniques.} Various detectors~have been proposed to detect malicious packages in the NPM~ecosystem, including rule-based~\cite{huang2024spiderscan,li2023malwukong,zheng2024OSCAR,Pohl2024runtimeProtect,GuardDog,duan2020maloss}, learning-based \cite{yu2024maltracker,nguyen2024classifyingByDynamic,Halder2024metainfo,huang2024donapi,zhang2024cerebro,ladisa2023crossLanguage,sejfia2022amalfi,wang2025Malpacdetector}, and LLM-based techniques~\cite{zahan2024detectionByLLM, guo2026bridging}. Rule-based detectors often~use static and/or dynamic analysis to model malicious behavior as patterns, and match them against predefined rules. 
Learning-based detectors typically adopt static and/or dynamic analysis to extract features, and train classifiers for detection. Wang et al.~\cite{wang2025Malpacdetector} further use an LLM to summarize malicious behavior features before training a conventional classifier. \modified{In both cases, the two analyses are usually employed in parallel (e.g., static and dynamic~features used independently) or with dynamic analysis serving to confirm static results.} 
LLM-based detectors leverage LLMs to directly understand the malicious behavior through prompt engineering, but are constrained in handling complex program behavior.

\tosem{However, existing detectors suffer from inadequate modeling of program behavior and limited interpretability, which hinder both their effectiveness and their practical adoption. Specifically, we identify the following four main limitations. The first three limitations (\textbf{L1}, \textbf{L2} and \textbf{L3}) concern how program behavior is modeled and represented, while the fourth (\textbf{L4}) limitation concerns how the resulting representation is reasoned about to reach a verdict.}

\textbf{L1: Inadequate Behavior Modeling of Obfuscated Code.} 
Several detectors~\cite{ladisa2023crossLanguage,sejfia2022amalfi} treat obfuscation as a feature~in~the learning process, resulting in~false positives, as such patterns~are~also present in benign packages. Huang et al.~\cite{huang2024spiderscan} designed~a~dedicated classifier to directly~detect obfuscation, and treated it as a sign of maliciousness, which leads to false positives when benign~packages are obfuscated. Instead, after detecting~obfuscation, Huang et al.~\cite{huang2024donapi} adopt dynamic analysis to model~behavior sequences, but noise in such sequences can still cause false positives. Zahan et al.~\cite{zahan2024detectionByLLM} use an LLM to identify the presence of~obfuscation, and incorporate it as one of the factors in their decision-making process. While this allows for certain flexibility,~it heavily relies on the capability of the LLM to discern malicious patterns. Some detectors~\cite{zhang2024cerebro,yu2024maltracker} completely ignore obfuscated code, potentially leading to false negatives.


\textbf{L2: Ignoring Object-Centric Features of JavaScript.}~{Detectors~\cite{yu2024maltracker,huang2024spiderscan,huang2024donapi,zhang2024cerebro,sejfia2022amalfi}} that rely on lexical or syntactic~patterns often struggle with JavaScript’s object-centric~nature, and thus fail to model the suspicious behaviors embedded within object-based constructs, which leads to false negatives. While program dependency graph (PDG)-based analyses~\cite{huang2024spiderscan,yu2024maltracker}~are effective at modeling program behavior by capturing control and data dependencies between statements, they exhibit weaknesses when analyzing object-oriented operations~(e.g., property access and object binding) as they are object-insensitive. This limitation results in missed detection of sensitive API calls. It creates opportunities for malicious packages to conceal harmful behaviors beyond the reach of existing detectors.


\textbf{L3: Lack of Synergy between Static and Dynamic~Analysis.} Existing detectors struggle to model behaviors that involve dynamic features, e.g., on-the-fly module loading~and~computed property access, which often leads to false negatives.~Dynamic analysis is crucial for capturing JavaScript's runtime~behavior. However, existing detectors either overlook dynamic analysis \cite{yu2024maltracker,zhang2024cerebro,zahan2024detectionByLLM}, or separately use it as confirmation after static~analysis~\cite{huang2024donapi,huang2024spiderscan} (e.g., detecting obfuscation or sensitive API calls that need further confirmation). Detectors that run static and dynamic analyses in parallel produce disjoint results without merging their evidence~\cite{duan2020maloss}. Purely~dynamic~detectors,~on the other hand, generate API call sequences, which are~often~overwhelmed by noise and cause false positives~\cite{zheng2024OSCAR}.~Consequently, no detector produces a unified representation that integrates dynamic evidence into static evidence, preventing static and~dynamic analysis from fully leveraging each other's strengths.


\textbf{L4: Semantic Information Loss in Behavior Abstraction and Lack of Interpretability.} \tosem{To make detection tractable, existing detectors do not reason over source code directly, but first abstract it into an intermediate representation, e.g., discrete lexical/syntactic or metadata features~\cite{yu2024maltracker,sejfia2022amalfi,ladisa2023crossLanguage}, normalized function or API call sequences~\cite{zhang2024cerebro,huang2024donapi,liang2025ClusteringInstallationScripts}, or behavior and call graphs~\cite{huang2024spiderscan}, including our previous work~\cite{huang2025ProfMal}. This abstraction inevitably obscures source-level semantics, and overlooks the surrounding code context, which prevents the detector from reasoning over code the way a human auditor would by directly reading it. Moreover, the decision stage typically emits only a package-level outcome, e.g., a binary label~\cite{yu2024maltracker,zhang2024cerebro} or an outlier flag from clustering~\cite{liang2025ClusteringInstallationScripts}, making it difficult for users to infer the underlying cause of the alarm or localize the responsible code. Some learning-based detectors partially mitigate this interpretability gap. \textsc{MalPacDetector}~\cite{wang2025Malpacdetector} uses an LLM to summarize malicious behavior features, and trains a conventional classifier on them; by mapping the resulting features to AST nodes, it provides line-level hints for suspicious code, but remains limited in explaining behaviors that span multiple functions or files. LLM-based detectors take a further step; e.g., \textsc{SocketAI}~\cite{zahan2024detectionByLLM} lets an LLM read the source code directly, thereby avoiding the semantic loss above. However, it feeds entire files to the LLM, which dilutes the relevant evidence among irrelevant code and inflates inference cost. Taken together, existing detectors still lack a mechanism that can preserve source-level semantics, focus analysis on security-relevant code, and provide fine-grained explanations of malicious behaviors across functions and files.}

\textbf{Our Approach.} 
\tosem{To address these limitations, we propose \tool, a detector that couples a unified behavior graph with a coordinated multi-agent framework for malicious NPM package detection. This work is a substantial extended version of our previous work~\cite{huang2025ProfMal}. \tool still focuses on malicious behaviors embedded during installation and import time, as the majority of malicious packages inject code at these stages~\cite{ohm2020backstabber}. The pipeline starts with a script analyzer that checks installation scripts for malicious shell commands and identifies entry files, providing the analysis scope for subsequent graph construction. The modeling limitations \textbf{L1}, \textbf{L2} and \textbf{L3} are addressed by the behavior graph construction introduced in our previous work~\cite{huang2025ProfMal}; i.e., \tool constructs behavior graphs that model control flows, control dependencies, and data dependencies between statements even in the presence of obfuscation~(\textbf{L1}); an object-sensitive static analysis resolves object references and aliasing for accurate identification of sensitive API calls and third-party library calls~(\textbf{L2}); and statically unresolved calls are resolved via dynamic analysis whose results are merged back into the behavior graphs, yielding a unified representation~(\textbf{L3}). We retain this behavior graph construction as the analysis substrate, and extend it in the following aspects. First, we refine the graph construction to analyze only the package's own code rather than downloading and parsing all third-party dependencies. Second, starting from suspicious nodes as anchors, a code slicer traverses control-dependence and data-dependence edges to extract multiple suspicious code slices from the behavior graph; each slice preserves the original source code while adding inline annotations derived from static analysis. Third, we introduce a multi-source third-party call characterization scheme; i.e., static analysis identifies third-party call sites, an LLM-based third-party enrichment agent consults registry metadata to inject module- and method-level semantics into the corresponding slices, and dynamic analysis captures the runtime API behavior sequences triggered behind each third-party call. Fourth, we replace the trained graph classifier with a coordinated multi-agent reasoning framework. A local behavior judge agent first reasons locally about each annotated slice, and a global behavior judge agent then follows the slices' CFG order to integrate all slice-level results and inter-slice relations into a comprehensive judgement. When this judgement remains undetermined, the global behavior judge agent identifies suspicious nodes whose evidence needs promotion, and a router agent invokes third-party enrichment or the behavior graph dynamic augmentor for those nodes before local and global reasoning are repeated. The framework further uses accumulated context and self-consistency verification to improve robustness. Fifth, we add a localization agent that pinpoints the malicious code snippets with explanations. Taken together, these extensions mark a paradigm shift from end-to-end graph learning to semantic-level evidence reasoning, thereby addressing \textbf{L4}.}

\textbf{Evaluation.}  \tosem{We}~compared \tool to five state-of-the-art detectors, \textsc{GuardDog}~\cite{GuardDog}, \textsc{Cerebro}~\cite{zhang2024cerebro}, \tosem{\textsc{ProfMal}~\cite{huang2025ProfMal}, \textsc{Malpacdetector}~\cite{wang2025Malpacdetector}}, and \textsc{SocketAI}~\cite{zahan2024detectionByLLM}. \tool achieved~the highest F1-score of \todo{98.1\%}, outperforming state-of-the-art detectors by \todo{3.5\% to 52.6\%}. \tosem{In terms of interpretability, \tool localized malicious code with a line-level F1-score of \todo{88.9\%}, and produced high-quality explanations for \todo{86.9\%} of the sampled malicious packages.} We ran all detectors \tosem{(together with \textsc{EMPHunter}~\cite{liang2025ClusteringInstallationScripts})} in a real-world setting and~monitored newly published NPM packages~for three~months. \tool detected \todo{597} previously unknown malicious packages,~achieving the lowest false positive rate of \todo{16.5\%}. All~these malicious packages have been confirmed and removed from~NPM.
\tosem{This evaluation also extends the evaluation in our previous work~\cite{huang2025ProfMal} by updating the compared detectors with recent state-of-the-art detectors, adding an LLM backbone selection experiment, redesigning the ablation study according to the extended pipeline, introducing an interpretability evaluation, and re-conducting the real-world detection.}

\tosem{
\textbf{Contribution.} In summary, this paper makes the following contributions.
\begin{itemize}[leftmargin=*]
    \item We propose \tool, a malicious NPM package detector that combines script analysis, unified behavior graph construction, and code slicing to extract multiple suspicious code slices augmented with static, third-party, and runtime evidence, preserving source-level semantics for downstream reasoning.
    \item We design a coordinated multi-agent reasoning framework that performs local judgement over individual slices, adaptively enriches uncertain evidence, and performs global judgement over slice-level results and inter-slice relations to improve detection accuracy and interpretability.
    \item We introduce a localization agent that maps the synthesized evidence back to concrete files and code slices, enabling \tool to explain whether a package is malicious and where the malicious behavior is implemented.
    \item We conducted experiments to demonstrate the effectiveness and practical usefulness of \tool. \tool achieved the highest F1-score among five state-of-the-art detectors, and detected \todo{597} previously unknown malicious packages during three months of real-world monitoring.
    \item \modified{We released the code of \tool at our website~\cite{tool_website}.}
\end{itemize}
}

\section{Motivating Examples}\label{sec:motivations}

\begin{figure}[!t]
	\centering
	\includegraphics[scale=0.89]{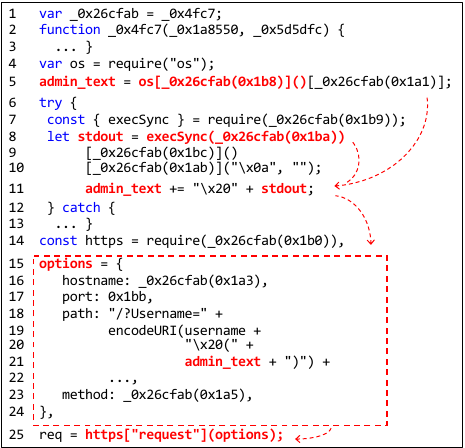}
	\vspace{-10pt}
	\caption{Code Snippet of \textit{bitsoex\_react-design-system\_14.1.4}}
	\label{fig:code_listing_2}
\end{figure}

Fig.~\ref{fig:code_listing_2} shows the code snippet of a malicious package~\textit{bitsoex\_react-design-system\_14.1.4}, illustrating how an attacker employs code obfuscation to conceal sensitive API calls and exfiltrate personally identifiable information (PII) via~network requests. Specifically, the attacker adopts syntactic symbol~obfuscation and dynamic property generation, which are common tactics used to evade detection \cite{ohm2020backstabber}. This code snippet defines~a calculation function \texttt{\_0x4fc7} at Line 2, which maps numeric codes to string literals at runtime, and binds it to an obfuscated variable \texttt{\_0x26cfab} at Line 1. It loads the module \texttt{os}~at~Line 4, and calls \texttt{os.userInfo()} at Line 5, where the property name \texttt{userInfo} is dynamically generated by \texttt{\_0x26cfab}. At Line 7, it uses \texttt{\_0x26cfab} to require \texttt{child\_process} and obtain \texttt{execSync}, then executes an obfuscated system command at Line 8 and appends the result to \texttt{admin\_text} at Line 11. Finally, it calls \texttt{https.request()} with its module name dynamically generated at Line 14, and sends an HTTPS request that embeds the stolen PII in the URL.~While~static analysis struggles to detect these obfuscated sensitive API~calls due to their dynamic nature (illustrating~\textbf{L1}), {it remains~capable of constructing control flows and data dependencies among these calls}. To resolve these obfuscated API calls, dynamic analysis is required to capture the underlying API invocations~\cite{zheng2024OSCAR,huang2024donapi}. This example emphasizes the importance of complementing static analysis with dynamic analysis, illustrating \textbf{L3}.

\begin{figure}[!t]
	\centering
	\includegraphics[scale=0.9]{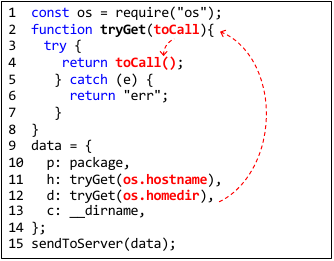}
	\vspace{-10pt}
	\caption{Code Snippet of \textit{automation.samples\_0.1.15}}
	\label{fig:code_listing_1}
\end{figure}

Fig.~\ref{fig:code_listing_1} shows the code snippet of a malicious package~\textit{automation.samples\_0.1.15}, illustrating how an attacker leverages JavaScript's object-centric features, specifically function object reference and aliasing, to hide sensitive API calls and exfiltrate PII. Specifically, the attacker builds a \texttt{data} dictionary at Line 9, containing PII by passing the {function object} of sensitive APIs \texttt{os.hostname} at Line 11 and \texttt{os.homedir} at Line 12 to the function \texttt{tryGet} defined at Line 2. The function \texttt{tryGet} then invokes these APIs via \texttt{toCall()} at Line 4, which acts as an alias for the actual API calls \texttt{os.hostname()} and \texttt{os.homedir()}. This indirection via function object references and aliasing not only hides the direct API calls but also reflects JavaScript's object-centric features, where in this example, functions are treated as objects and passed around. This tactic weakens detectors~\cite{huang2024spiderscan,zhang2024cerebro,yu2024maltracker,sejfia2022amalfi} that rely on direct identifier resolution, illustrating \textbf{L2}. This~example emphasizes the need for object-sensitive analysis to detect sensitive API calls and thus detect malicious behavior.

\begin{figure}[!t]
	\centering
	\includegraphics[scale=0.9]{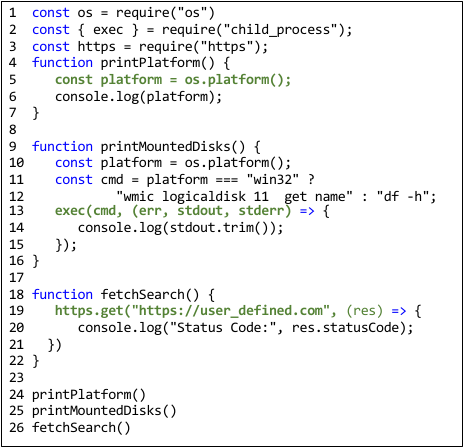}
	\vspace{-10pt}
	\caption{Code Snippet of \textit{echo-tool-1.0.1}}
	\label{fig:code_listing_3}
\end{figure}

Fig.~\ref{fig:code_listing_3} shows the code snippet of a benign package \emph{echo-tool-1.0.1}, which performs a simple startup check, i.e.,~printing platform information, listing mounted disks, and testing~network connectivity. Specifically, it detects the operating system using \texttt{os.platform()} at Line 5, and logs the result at~Line 6. It then uses \texttt{child\_process.exec()} at Line 13 to execute a platform-specific shell command, determined by the value of \texttt{platform}. Finally, it tests network connectivity via \texttt{https.get()} at Line 19. From a static analysis perspective, there is no direct data dependency among \texttt{os.platform()} (Line 5), \texttt{child\_process.exec()}, and \texttt{https.get()}. Each API is invoked independently, and no sensitive data~is transferred or leaked. Hence, a static detector can identify~this behavior as benign. However, from a dynamic analysis~perspective that relies on behavior sequence, the series of operations, i.e., system inspection via \texttt{os.platform()}, command execution via \texttt{child\_process.exec()}, and network communication via \texttt{https.get()} is similar~to~the~behavioral pattern shown in Fig.~\ref{fig:code_listing_1}, which indicates PII leakage. {Without the code structure and context specifying the relations between these API calls, this similarity leads to false positives~in~purely dynamic or behavior-sequence-based detectors~\cite{huang2024donapi,zheng2024OSCAR}}. This example also emphasizes the critical need for complementing dynamic analysis with static analysis, illustrating~\textbf{L3}.

\tosem{These examples show that static and dynamic program analysis are necessary for modeling program behaviors, including obfuscated calls, object-centric aliases, and runtime traces. However, behavior modeling alone is insufficient for reliable semantic reasoning. Existing detectors often abstract source code into intermediate representations, including behavior graphs. While these abstractions can preserve control and data dependencies, they inevitably hide part of the source-level evidence needed for reasoning, especially for LLM-based reasoners that rely on code context and concrete expressions to infer intent. In Fig.~\ref{fig:code_listing_2}, for example, graph analysis can recover that an obfuscated call resolves to \texttt{execSync}, that its return value flows into \texttt{admin\_text}, and that the collected value later reaches an \texttt{https.request}. Such graph nodes and edges capture the existence of sensitive calls and data flows, but they compress the original statements into abstract facts; i.e., a node may only indicate that an HTTPS request is issued, while the concrete~expression that dynamically assembles the URL and the exact source-level context showing how PII is embedded into the request are no longer directly visible to the reasoner. A natural alternative is to reason over the source code itself, which is closer to how human auditors inspect malicious behavior. Yet directly analyzing raw code is also unreliable, especially for obfuscated code; i.e., identifiers such as \texttt{\_0x26cfab} and dynamically generated properties can mislead a reasoner into incorrect interpretations, while irrelevant statements may dilute the security-critical logic. Therefore, reliable reasoning requires a source-code-level representation that is both focused on security-relevant statements and guided by facts recovered through program analysis; i.e., static analysis can identify sensitive API calls, recover aliases and~object~references, while dynamic analysis can resolve runtime-generated calls and values, expose concrete API behavior sequences, and attach these observations back to the corresponding code. Together, these observations highlight the need for faithful code views that preserve source-level context while being enriched by static and dynamic analysis, illustrating \textbf{L4}.}
	
\section{Approach}

\begin{figure*}[!t]
	\centering
	\includegraphics[scale=0.37]{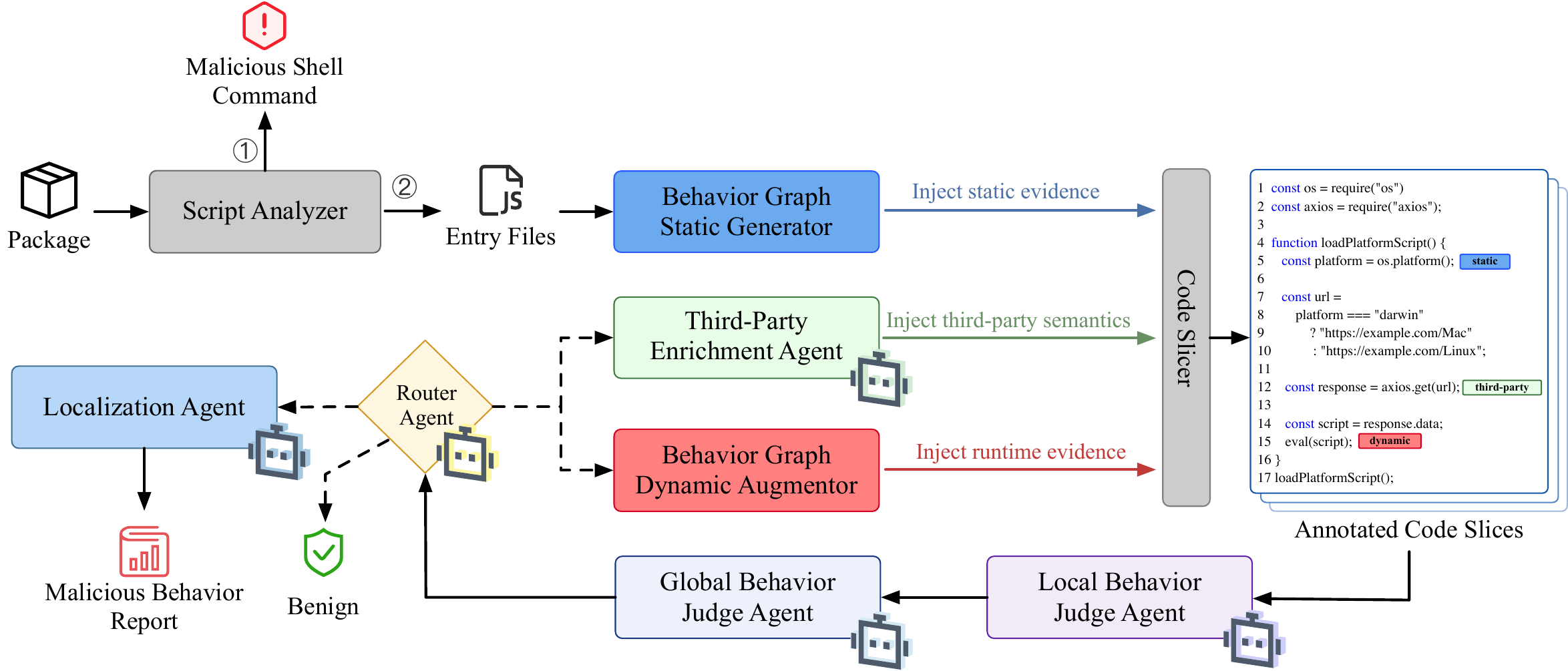}
	\vspace{-5pt}
 	\caption{Approach Overview of \tool}
    \label{fig:overview}
\end{figure*}

\tosem{To address the limitations of existing detectors, we propose \tool, a unified graph-based approach that leverages the synergy between static and dynamic analysis and coordinated LLM agents to detect malicious NPM packages. As shown in Fig.~\ref{fig:overview}, \tool consists of the following key modules.}

\begin{itemize}[leftmargin=*] 
\item \textbf{Script Analyzer} (Sec.~\ref{sec:scripts_analyzer}). This module scans a package's installation scripts to detect malicious shell commands. It also identifies all entry files used during installation and import time, laying the foundation for subsequent behavior graph construction and agent reasoning.

\item \textbf{Behavior Graph Static Generator} (Sec.~\ref{sec:behavior_graph_static}). Using the script analyzer's outputs, this module constructs behavior graphs by modeling control flows, control dependencies, and data dependencies~between statements. \tosem{It further identifies sensitive API calls, third-party library calls, and statically unresolved calls through object-sensitive analysis, thereby addressing limitations \textbf{L1} and~\textbf{L2}.}

\item \textbf{Code Slicer} (Sec.~\ref{sec:code_slicer}). Starting from suspicious nodes in the statically generated behavior graph, this module first extracts multiple independent suspicious code slices, and converts graph-level facts into source-level views for downstream reasoning. \tosem{Static and dynamic evidence is integrated into slices through behavior graph node attributes; and third-party evidence is integrated into slices directly from the third-party enrichment agent's output. Each slice preserves the original source code while carrying inline annotations, retaining source-level semantics and focused security-relevant context for subsequent LLM-based reasoning, thereby addressing the semantic-loss aspect of \textbf{L4}.}

\item \textbf{Local Behavior Judge Agent} (Sec.~\ref{subsub:local_judge}). This agent performs local judgement over each annotated code slice and produces a slice-level verdict with supporting evidence.

\item \textbf{Global Behavior Judge Agent} (Sec.~\ref{subsub:global_judge}). This agent integrates all slice-level results and inter-slice relations according to the slices' CFG order to derive a comprehensive entry-level judgement.

\item \textbf{Router Agent} (Sec.~\ref{subsub:router}). When the current entry-level judgement remains undetermined, this agent identifies suspicious nodes whose evidence needs promotion, and decides whether to request third-party semantic enrichment or dynamic augmentation.

\item \textbf{Third-Party Enrichment Agent} (Sec.~\ref{sec:third_party_enrichment}). When selected by the router agent, this module retrieves module- and method-level semantics from registry metadata, and injects them into the corresponding annotated code slices. \tosem{The enrichment agent returns the recovered semantics to the code slicer, which injects them by node identifier, helping the local behavior judge agent reason about library calls that are opaque from local source code alone.}

\item \textbf{Behavior Graph Dynamic Augmentor} (Sec.~\ref{sec:behavior_graph_dynamic}). When selected by the router agent, this module executes the package in a sandbox, and augments the behavior graphs with dynamic evidence for the routed suspicious nodes, addressing limitations \textbf{L1} and \textbf{L3}. \tosem{It is triggered only for specific node types, namely \emph{conditional}, \emph{unresolved}, \emph{eval}, and \emph{third-party} nodes whose semantics failed to be resolved by third-party enrichment.}

\item \textbf{Localization Agent} (Sec.~\ref{sec:localization}). Once \tool flags a package as malicious, this agent maps the synthesized evidence back to concrete files and code slices, pinpointing the malicious code snippets with explanations. \tosem{This addresses the interpretability aspect of \textbf{L4}.}
\end{itemize}


\subsection{Script Analyzer}\label{sec:scripts_analyzer}

Our script analyzer processes the \textit{package.json} file~of~a~package to (1) detect malicious shell commands in installation scripts, (2) identify entry files used during installation and import time, \tosem{and (3) parse dependency manifest}.

\textbf{Malicious Shell Command Detection.}
During package~installation via \texttt{npm install}, the package manager~automatically executes shell commands specified in the installation script fields of \textit{package.json}, e.g., \texttt{preinstall}, \texttt{install},~and \texttt{postinstall}. These scripts are executed without user~interaction, and can contain arbitrary shell commands, thereby introducing a significant attack surface, as attackers can~leverage them to run malicious code during installation. To address this threat, we first detect potential malicious shell commands in these fields. If a malicious shell command is detected, \tool reports it while continuing to analyze~the~package code for any further malicious behavior.

We detect malicious shell commands based on their behavior, similar to the malicious shell command~detection approach used in \textsc{SpiderScan}~\cite{huang2024spiderscan} and \textsc{SocketAI} \cite{zahan2024detectionByLLM}. \tool prompts the LLM to analyze the behavior~of the command and judge its maliciousness. Following prior work, our prompt design adopts the structure of \textsc{SpiderScan}, which consists of role specification, task description, behavior guidelines, and a structured output format. We extend this structure in three aspects. First, the prompt classifies each command into \texttt{benign}, \texttt{warning}, or \texttt{malicious}, and requires the LLM to assign \texttt{malicious} only when the evidence is conclusive. Second, it uses an expanded taxonomy of security-relevant behaviors distilled from prior studies~\cite{zahan2024detectionByLLM,huang2024donapi,huang2024spiderscan,ohm2020backstabber,zhou2024largeScaleAnalysis,trizna2021shell}, including data exfiltration, suspicious script or binary execution, dropper behavior, download-and-execute behavior, critical file tampering, process injection, obfuscation and encoding, reverse shells, system shutdown, critical file deletion, unusual URL interaction, non-typical Node execution, and resource exhaustion. \tosem{Third, the script analyzer supports an optional \texttt{read\_files} action for local shell-style scripts invoked by the command, while explicitly excluding JavaScript files from this reading step because they are analyzed by the downstream pipeline.} 
\tosem{Specifically, some installation commands delegate their behavior to locally provided shell-style scripts, mainly including POSIX shell scripts (e.g., \texttt{.sh}, \texttt{.bash}, and \texttt{.zsh}) and PowerShell scripts (e.g., \texttt{.ps1}). When the command invokes such a script, \tool allows the LLM to issue a bounded~\texttt{read\_files} request for the script content, so that the final judgement reflects the complete install-time command chain rather than the command alone. This reading step is restricted to local shell-style scripts inside the package and excludes JavaScript files.}
The prompt snippet used for malicious shell command detection is presented in Fig.~\ref{fig:shell_command_template}, while the complete prompt is available at our replication website~\cite{tool_website}.

\begin{figure}[!t]
	\begin{tcolorbox}[
		title=Prompt Snippet,
		colback=white,
		colframe=black,
		fonttitle=\bfseries\footnotesize,
		boxrule=0.5pt,
		arc=2pt,
		left=1pt, right=1pt, top=1pt, bottom=1pt,
	]
	\footnotesize
	\tosem{\textbf{Task:} You are a security expert with extensive experience in shell programming and security risk assessment across common operating systems. You will receive one input: \texttt{<Shell Command>}, which is executed in the \texttt{scripts} field of \textit{package.json}. Your task is to classify the command as \texttt{benign}, \texttt{warning}, or \texttt{malicious}. Only classify a command as \texttt{malicious} when the evidence is conclusive.}\\

	\textbf{Guidelines:} Malicious behaviors are defined as follows (with examples):
    1. Data Exfiltration: Commands that retrieve local data (e.g., passwords, system logs, /etc/passwd) and send it externally (e.g., via remote server, URL, or DNS lookup).\\
	\tosem{2. Binary Execution: Commands that execute binary files, especially from unknown or suspicious sources.}\\
	3. Download and Execution: Commands that both download and execute files in a single step.\\
    \ldots\\

	\tosem{\textbf{Reading Inner Shell Scripts:} If the command invokes a locally provided shell-style script, e.g., \texttt{bash setup.sh}, \texttt{sh tools/run.sh}, \texttt{pwsh install.ps1}, or \texttt{cmd /c build.bat}, request the script through \texttt{read\_files} and inspect its content before assigning the final label, so that the decision reflects both the command and the script behavior. Do not request JavaScript files.}\\
	
	\tosem{\textbf{Output:} Return a JSON object with either \texttt{\{"action": "read\_files", "paths": ["<path>"], "reason": "<reason>"\}} or \texttt{\{"action": "final", "label": "benign|warning|malicious", "explanation": "<explanation>"\}}.}
	\end{tcolorbox}
	\vspace{-10pt}
	\caption{Prompt Snippet for Malicious Shell Command Detection}
	\label{fig:shell_command_template}
\end{figure}

\textbf{Entry Files Extraction.}
To detect malicious behavior during import time, where harmful code is often embedded,~we~extract entry files, including JavaScript files referenced in~the \texttt{main}, \texttt{exports}, and \texttt{bin} fields of \textit{package.json}, which are typically triggered on \texttt{require} statements. Besides, we also use regular expressions to extract JavaScript files executed~in~the~installation scripts. The output of this step is a set of entry files.

\textbf{Dependency Manifest Parsing.}
To support the identification of third-party calls in the subsequent procedure, we parse the \texttt{dependencies} field of \textit{package.json}, and collect the package names together with their declared versions. \tosem{The extracted dependency metadata is used only by the behavior graph static generator (Sec.~\ref{sec:behavior_graph_static}) to recognize third-party call sites; semantic enrichment of those calls is performed later on demand by the third-party enrichment agent (Sec.~\ref{sec:third_party_enrichment}).}

\subsection{Behavior Graph Static Generator} \label{sec:behavior_graph_static}

Our static generator takes entry files \tosem{and third-party dependency metadata} as inputs, and constructs a behavior graph (BG) for each entry file based on the syntactic structure of the code. The generated BGs capture the overall behavior of the package, including its interactions with third-party libraries. 

\subsubsection{Overall Procedure} \label{subsec: overall_procedure}
To generate the behavior graph (BG), we first construct a code property graph (CPG) \tosem{for the package code} using Joern~\cite{Joern}. The CPG comprises the control flow graph (CFG), the control dependency graph (CDG), and the data dependency graph (DDG). The DDG models data dependencies within each single function, where nodes represent statements and edges capture the data dependencies between them. The CFG captures the control flows among these statements. The CDG captures control dependencies among statements, indicating which statements are guarded by control predicates. To unify control flows, control dependencies, and data dependencies, we integrate the CFG and CDG into the DDG by adding control flow and control dependency edges among~statements, resulting in a graph denoted as {DDG}$^{+}$. In parallel, we generate a call graph (CG) using Jelly~\cite{Jelly}, where call edges link call sites to their corresponding callees, enabling inter-procedural analysis across functions.

Next, we model each entry file as an implicit \textit{main} function, encompassing all code in the file's global scope. Starting from this implicit \textit{main} function of each entry file, we traverse~the nodes in {its} {DDG}$^{+}$ following the control flow~to~mimic~the~execution, and use the CG to step into another function at a function call node. The traversal runs {object-sensitive}~and~flow-sensitive analysis by analyzing each expression within a statement. \modified{For example, when encountering the statement \texttt{tryGet(os.hostname)}, the analysis first resolves the subexpression \texttt{os.hostname} and then proceeds to the outer call expression \texttt{tryGet} following AST semantics.}

Sensitive API calls are important to model malicious~behavior. One major challenge in identifying sensitive API calls lies in the inaccuracy caused by indirect calls as shown~in~{Section}~\ref{sec:motivations}, where the actual target of an API call is hidden by~an~intermediate variable. To address this, the goal~of our analysis is to identify all API calls and then match them against a predefined list of sensitive APIs to pinpoint the sensitive ones. Specifically, we track object references (e.g., resolving the property access \texttt{os.hostname} to its corresponding function object, as shown in Fig.~\ref{fig:code_listing_1}), and maintain alias~relations~(e.g.,~treating~the~identifier \texttt{toCall} in \texttt{tryGet()} as an alias for the object referenced by \texttt{os.hostname}, as shown in Fig.~\ref{fig:code_listing_1}) throughout the code. 

Additionally, \tosem{our analysis identifies third-party API call sites when the imported module matches the dependency metadata and the invoked method path can be statically determined, e.g., recognizing \texttt{axios.post} as a call to module \texttt{axios} and method \texttt{post}.} Our analysis also detects call sites that cannot~be resolved statically, such as unidentified sensitive API~calls~(e.g., \texttt{os[0x26cfab(0x1b8)]()} in Fig.~\ref{fig:code_listing_2}, where the property value is dynamically generated), or unresolved calls whose callees are dynamically bound. These unresolved, conditional, and third-party call sites may later be resolved on demand when the router agent (Sec.~\ref{subsub:router}) selects third-party enrichment (Sec.~\ref{sec:third_party_enrichment}) or dynamic augmentation (Sec.~\ref{sec:behavior_graph_dynamic}).

After analyzing each node in {DDG}$^{+}$, we add it to the BG and connect it to related nodes based on their relations in {DDG}$^{+}$. For function call nodes, we attach analysis-derived~attributes. Formally, the resulting BG is denoted as a directed graph~$G = $ $(V, E)$,
 {where} $V$ is the set of nodes, and each node represents a statement; and $E$ is the set of edges connecting nodes, including control-flow and data-dependency edges. Each function call node $v \in V$ is annotated with $\langle T_v,C_v,U_v,E_v,P_v \rangle$.
Here, {$T_v = 1$ denotes a \emph{sensitive} node, indicating that the node represents a sensitive API call. $C_v = 1$ denotes a \emph{conditional} node, indicating that the node is potentially sensitive~but~requires further assessment. For example, \texttt{fs.readSync} is considered sensitive due to its potential to leak data, but the actual risk depends on the usage context, such as its arguments and the data being accessed. 
$U_v = 1$ denotes an \emph{unknown} node, meaning the call cannot be resolved statically. $E_v = 1$ indicates an \emph{eval} node, which executes dynamically generated code via \texttt{eval}. \tosem{$P_v = 1$ denotes a \emph{third-party} node.}}



To track object references and maintain alias relations, we mainly analyze four types of statements, i.e., \textbf{Require}, \textbf{Property Access}, \textbf{Assignment}, \textbf{Function Call}, during the traversal. These four types of statements are central to how~objects are created, propagated, and aliased throughout the program.

The traversal is single-pass. As we traverse the {DDG}$^{+}$, we maintain two mappings at each program point.
\begin{itemize}[leftmargin=*]
    \item \textbf{Identifier Map} ($\mathit{idMap}$): {Its keys are identifiers $id$, and~its values are the corresponding objects $obj$, allowing the~retrieval of the object associated with a given identifier by $obj =$ $\mathit{idMap}[id]$. It maintains alias relations by recording the object each identifier refers to during the mimicked execution, enabling us to track how different identifiers point to the same object across the program. Since identifier visibility varies across different scopes, $\mathit{idMap}$ also maintains information about the scope in which each identifier is accessible. To capture these visibility relations, it employs a stack-based structure, updating the available identifiers as the traversal enters or exits code scopes (e.g., function entries and exits). Specifically, $\mathit{idMap}$ is updated upon assignment statements where the left-hand side is a single identifier (Sec.~\ref{subsub:assignment}).}

    \item \textbf{Property Map} ($\mathit{propMap}$): {Its keys are object-property~pairs $\langle obj, prop \rangle$ (denoting the property $prop$ of an object $obj$), and its values are the corresponding target~objects, allowing the retrieval of the object referenced by a given object-property pair by $obj_{ref} = \mathit{propMap}[\langle obj, prop \rangle]$. It tracks the object referenced by each property of an object, thereby facilitating the analysis of property-based references. Specifically, $\mathit{propMap}$ is updated upon assignment to the property of an object (Sec.~\ref{subsub:assignment}).}
\end{itemize}

For each object, we maintain two attributes.
\begin{itemize}[leftmargin=*]
    \item \textbf{Qualified Object Path ($qPath$)}: It records the hierarchical property access sequence leading to the first definition of the object itself. The $qPath$ is denoted as:
    \[  
        \begin{aligned}
            qPath = \langle root, props\rangle, \; \text{with} \; props = [p_1, ..., p_n] 
        \end{aligned}
    \] 
    {where $root$ denotes the original root object, which~is~the~initial object in the chain of property accesses, and $props$~is~the list of properties accessed in order; e.g., the function object of \texttt{os.userInfo} has a $qPath$ of $\langle obj_{os}, [userInfo] \rangle$}, where $obj_{os}$ is the object referred to by the identifier $os$.
    \item \textbf{Qualified Object Name ($qName$)}: It is a string that~represents the object's full qualified name, and is used~for~constructing the full qualified name of an API call; {e.g., the builtin object \texttt{fs} has a $qName$ of ``fs'',~while the function object \texttt{os.userInfo} has a $qName$ of ``os.userInfo''.}
\end{itemize}

Before our traversal, we pre-build an object for every builtin module~\cite{node_js_documentation} by initializing its $qName$ to the module’s name and its $qPath$ to $\langle obj_{builtin}, \phi \rangle$ where $obj_{builtin}$~is~the~builtin module object. These objects are stored in a set $\mathcal{B}$~for~subsequent analysis. This ensures that each builtin module corresponds to a unique object during our traversal. \tosem{We also take the third-party dependency metadata extracted by the script analyzer (Sec.~\ref{sec:scripts_analyzer}) as a set $\mathcal{D}$ of package names and declared versions. This metadata is used to distinguish dependency-matched third-party library roots from local modules during \texttt{require} analysis.}

\subsubsection{Require}\label{subsub:require}
The \texttt{require} {statement} is used to import builtin, local or external modules, and can be defined as:
\[
    \mathrm{require}(module)
\]
where $module$ is an expression representing the name or path of the module. This expression can be a string literal, variable, or computed value. \tosem{We handle it by the following three cases.}
\begin{enumerate}[leftmargin=*]
    \item \textbf{Builtin module.} 
    {If $module$ is a string literal and matches against a Node.js builtin module, we retrieve the corresponding pre-built object $obj_{builtin}$ from the set $\mathcal{B}$.}

    \item \tosem{\textbf{Third-party library.} If $module$ is a string literal and matches a package name in $\mathcal{D}$, or starts with a dependency package name followed by a path separator, we create or retrieve a third-party module object $obj_{tp}$, set its $qName$ to the matched package name, and initialize its $qPath$ as $\langle obj_{tp}, \phi \rangle$ for a package-root import or $\langle obj_{tp}, [p_1,\ldots,p_k] \rangle$ for a subpath import. The declared version from $\mathcal{D}$ is preserved as metadata of $obj_{tp}$.}

    \item \tosem{\textbf{Dynamically computed module.}} If $module$ is either a string literal not in the builtin list \tosem{and not matched in the third-party library case}, or~a~dynamically computed value (i.e., the module name is computed at runtime), we create a new module object $obj_{new}$, set its $qName$ to $\phi$, and initialize its $qPath$ as $\langle obj_{new}, \phi \rangle$.


\end{enumerate}

\tosem{By assigning qualified names to builtin modules and dependency-matched third-party libraries, we ensure accurate resolution of builtin API calls and recognize third-party API calls. For other modules, the qualified name is left empty ($\phi$), allowing the identification of unresolved call sites in subsequent static analysis. In all cases, the value returned by \texttt{require} is treated as an object and used in subsequent procedures.}

\subsubsection{Property Access}\label{subsub:property_access}
Property access plays a fundamental role in interacting with objects, and occurs primarily in two forms, e.g., {\texttt{a.b.c} and \texttt{a[b][c]}}. In both cases, \texttt{a}~is~the~leftmost identifier, while \texttt{b} and \texttt{c} represent literal property names. {When a property, such as \texttt{b} or \texttt{c} in \texttt{a[b][c]}, is not a literal, we use a wildcard symbol $\bot$ to denote the non-literal property.}

For every property access expression $e$, we represent it as {$e = \langle id_0, [p_1, \dots, p_n] \rangle$}, where $id_0$ is the leftmost identifier and $[p_1, \dots, p_n]$ is the left-to-right sequence of property names. {For example, the expression \texttt{a.b.c} is represented as $\langle a, [b, c] \rangle$, and \texttt{d[e]} as $\langle d, [e] \rangle$. To handle a nested expression, which~reflects multi-level property access, we decompose $e$ into a sequence of atomic steps:}
\[  
    \begin{aligned}
    \langle id_{0}, [p_{1},\dots ,p_{n}]\rangle \Rightarrow  \langle &id_{1}, [p_2,...,p_{n}] \rangle \Rightarrow  \dots \Rightarrow  \langle id_{n-1}, [p_{n}] \rangle,\\
    \text{with} \; &id_{j} = \langle id_{j-1}, p_j\rangle
    \end{aligned}
\]
{where $id_j$ is an identifier that refers to the object obtained~by accessing property $p_j$ of $id_{j-1}$, and is used as the base identifier in the next atomic step.}

For each atomic step $id_{j} = \langle id_{j-1}, p_j\rangle$ in property access, we operate as follows.
\begin{enumerate}[leftmargin=*]
    \item \textbf{Identifier to object mapping.}
        We retrieve the referenced object $obj_{j-1}$ of the identifier $id_{j-1}$ from $\mathit{idMap}$. 
    
    \item \textbf{Property lookup.} 
        We obtain the object referenced by the object-property pair $\langle obj_{j-1}, p_j \rangle$ from $\mathit{propMap}$. {If such a mapping exists, $obj_{pl}$ is the target object of the lookup.}
        
    \item \textbf{Qualified path initialization.}
        If the property lookup~fails to find an object of the given object-property pair, possibly because the object-property pair has not been~explicitly assigned before, such as originating from a builtin~module \tosem{or a third-party library} (e.g., \texttt{os.hostname}, whose function object is not {pre-assigned, i.e., no explicit object is created~for~the~property before}), or the property is $\bot$, we create a new object $obj_{new}$, and add it to the~map~by~$\mathit{propMap}[\langle obj_{j-1}, p_j \rangle] = obj_{new}$, which would not disrupt the remainder of the analysis based on that object. 
         For $obj_{new}$, we set its qualified path through  concatenation, i.e., $obj_{new}.qPath = \langle obj_{j-1}.qPath.root,$ $ obj_{j-1}.qPath.props \;\Vert\; p_j \rangle$, where $\Vert$ denotes appending property $p_j$ to the existing sequence $obj_{j-1}.qPath.props$.
         
    \item \textbf{Object aliasing.}
        {Let $obj_{res}$ denote the resulting object~for this property access. Specifically, $obj_{res} = obj_{pl}$~if~the~property lookup succeeds, or $obj_{res} = obj_{new}$ if the property lookup fails. Upon the assignment $id_j = obj_{res}$, we update the map by setting $\mathit{idMap}[id_j] = obj_{res}$ (Sec.~\ref{subsub:assignment}).}
\end{enumerate}
In each atomic step, we obtain the corresponding referenced object, and set one of its aliases to an identifier, which then~serves as the leftmost identifier in the un-nested expression. Once all atomic steps are complete, we obtain a resulting object as the result of the property access.

\subsubsection{Assignment}\label{subsub:assignment}
Assignment is a fundamental operation that establishes or updates the binding between variables, object properties, and values. We represent the assignment as:
\[
    expr_{lhs} = expr_{rhs}
\]
The analysis of an assignment proceeds in two phases:~1)~resolving the right-hand side (RHS) expression to an object, and
2) analyzing this object with the left-hand side (LHS) target, which may be an identifier or a property access expression.
\begin{enumerate}[leftmargin=*]
    \item \textbf{Right-hand side resolution.}
    {The goal is to resolve $expr_{rhs}$ to a specific object $obj_r$ based on the type of $expr_{rhs}$ as follows: (1) if $expr_{rhs}$ is a property access expression,~we analyze it to obtain the target object (Sec.~\ref{subsub:property_access}); (2) if it is an identifier, we retrieve the referenced object~via~$\mathit{idMap}$; (3) if it is a string literal, we create a new object, and~set its qualified name to the literal value; (4) if it is a \texttt{require}, we analyze it to obtain the corresponding object (Sec.~\ref{subsub:require}); and (5) if it is a function call, we analyze~it  to obtain the target object (Sec.~\ref{subsub:function_call}).}

    \item \textbf{Left-hand side binding.}
    Once $obj_r$ is determined,~we~analyze $expr_{lhs}$ as follows: (1) if $expr_{lhs}$ is a standalone identifier $id$, we set $\mathit{idMap}[id] = obj_r$, making $id$ an alias to $obj_r$; and (2) if $expr_{lhs}$ is a property access, we follow the property access procedure (Sec.~\ref{subsub:property_access}), leaving the last property unresolved to the form $\langle obj_{last}, p_{last} \rangle$, and then update the property map as $\mathit{propMap}[\langle obj_{last}, p_{last} \rangle] = obj_r$. {For example, in the assignment \texttt{a.b.c = r}, we resolve up to \texttt{a.b} and retain the last property \texttt{c}, and update the property map as $\mathit{propMap}[\langle obj_{a.b}, \text{c} \rangle] = obj_r$.}
 \end{enumerate}

\subsubsection{Function Call}\label{subsub:function_call} 
{Calls to user-defined functions (including third-party and custom functions) or builtin APIs are important program behaviors. A call expression is represented as:}
\[
    expr_{callee}(expr_{a_1},...,expr_{a_n})
\]
where {$expr_{callee}$ is the callee expression}, which can be~an~identifier, a property access expression, or \texttt{require}, etc. $expr_{a_i}$ is an argument expression. Each call node is analyzed as follows.
\begin{enumerate}[leftmargin=*]
\item \textbf{User-defined function call.} If the CG resolves the callee for $expr_{callee}$, {the node is a user-defined function call.} An inter-procedural control flow edge is added from the caller to the callee in the BG, and we traverse into the callee. If $expr_{callee}$ is not \texttt{require}, formal parameters are bound to actual arguments to track how argument objects propagate. 
{For each formal parameter $id_{fp_i}$, we record an alias to the object resolved from the corresponding argument expression $expr_{a_i}$, and update $\mathit{idMap}$ accordingly. During the traversal, return statements are handled in a flow-sensitive way. For each return expression $expr_r$, the object resolved from~it~is used as the function’s return value. If there is no return~value, a new object is created with its qualified name set to $\phi$.}

\begin{table}[!t]
    \centering
    \footnotesize
    \caption{Part of the Sensitive API List}
    \vspace{-10pt}
    \label{tab:sensitive_api_list}
      \begin{tabular}{ 
          >{\centering\arraybackslash}m{3.0cm} 
          >{\centering\arraybackslash}m{3.8cm}
          >{\centering\arraybackslash}m{0.7cm}
          }
      \toprule
      \textbf{Qualified Full Name}   & \textbf{Behavioral Type}  & \textbf{ARD}          \\
      \midrule
      os.hostname                    &   System Information Retrieval &    \ding{55}     \\
      zlib.brotliCompress            &   Data Compression             &    \ding{55}     \\
      child\_process.execSync        &   Command Execution            &    \ding{51}     \\
      net.Socket                     &   Network Creation             &    \ding{55}     \\
      https.get                      &   GET Request                  &    \ding{55}     \\
      fs.readSync                    &   File Reading                 &    \ding{51}     \\
      dns.resolve                    &   DNS resolution               &    \ding{55}     \\
      crypto.createCipheriv          &   Cipher Creation              &    \ding{55}     \\
      \bottomrule
    \end{tabular}
  \end{table}

\item \textbf{API call.} If the CG cannot resolve the callee, we compute the full name $\mathcal{F}$ of the call for API identification. Specifically, we resolve $expr_{callee}$ to obtain the corresponding object $obj_{callee}$ based on its type, e.g., by retrieving the referenced object from $\mathit{idMap}$ if it is an identifier, or by resolving the property chain if it is a property access. The full name is computed based on $obj_{callee}.qPath$.~In~detail, given $obj_{callee}.qPath = \langle o_{root}, [p_1,...,p_n] \rangle$, $ \mathcal{F} = o_{root}.qName + (.).concat([p_1,...,p_n])$. Then, we determine whether $\mathcal{F}$ appears in a predefined sensitive API list that is constructed by following prior work \cite{huang2024spiderscan, huang2024donapi, zhang2024cerebro, yu2024maltracker, zheng2024OSCAR}. A partial list is shown in Table~\ref{tab:sensitive_api_list}, and the full list is available at our website~\cite{tool_website}. Each entry in the list includes~the~API’s full name, behavioral category, and an ARD \modified{(Arguments \& Return Values Dependent)} flag indicating whether its behavior depends~on~usage~context,~i.e.,~arguments and return values. If $\mathcal{F}$ matches a sensitive API, the call node {$v$} is marked as \emph{sensitive} ($T_v = 1$). If the ARD flag is true, the node is marked as \emph{conditional} ($C_v = 1$). \tosem{If $o_{root}$ corresponds to a third-party module object derived from $\mathcal{D}$ and $\mathcal{F}$ can be computed, we mark the call node as \emph{third-party} ($P_v = 1$) and record it as a third-party API call. The recorded metadata contains the module name, declared version, and method path $[p_1,...,p_n]$; for example, for \texttt{axios.post}, we record \texttt{axios} as the module name and \texttt{post} as the method name.} Since this is an API call, we do not bind actual arguments to formal parameters. Instead, we record the argument values as part of the call’s metadata for subsequent analysis. Finally, a new object is created to represent the return value of the API call, and $\mathcal{F}$ is assigned as its qualified name.

\item \textbf{Unknown call.}
If the full name $\mathcal{F}$ cannot be statically~computed, either because $o_{root}.qName = \phi$ or $qPath$ contains a property that is $\bot$ (e.g., \texttt{\_ox26cfab(0x1b8)} in the \texttt{os[\_ox26cfab(0x1b8)]}), the node {$v$} is marked~as~\emph{unknown} ($U_v = 1$). Such unknown calls potentially indicate sensitive API calls or unresolved {user-defined function calls}. {In such cases, we create a new object with its qualified~name set to $\phi$ to represent the unresolved return value of the call, allowing subsequent dynamic analysis to properly resolve~it.}


\item \textbf{Eval.} If $expr_{callee}$ equals \texttt{eval} and its arguments are dynamically generated, we mark the node $v$ as \textit{eval} ($E_v = 1$). 
{We also create a new object for this call, with its qualified name set to $\phi$, representing the return value of \texttt{eval}.}
\end{enumerate}

After analyzing each function call node, the node along with its attributes is added to the BG.

\begin{figure}[!t]
  \centering
  \includegraphics[scale=0.388]{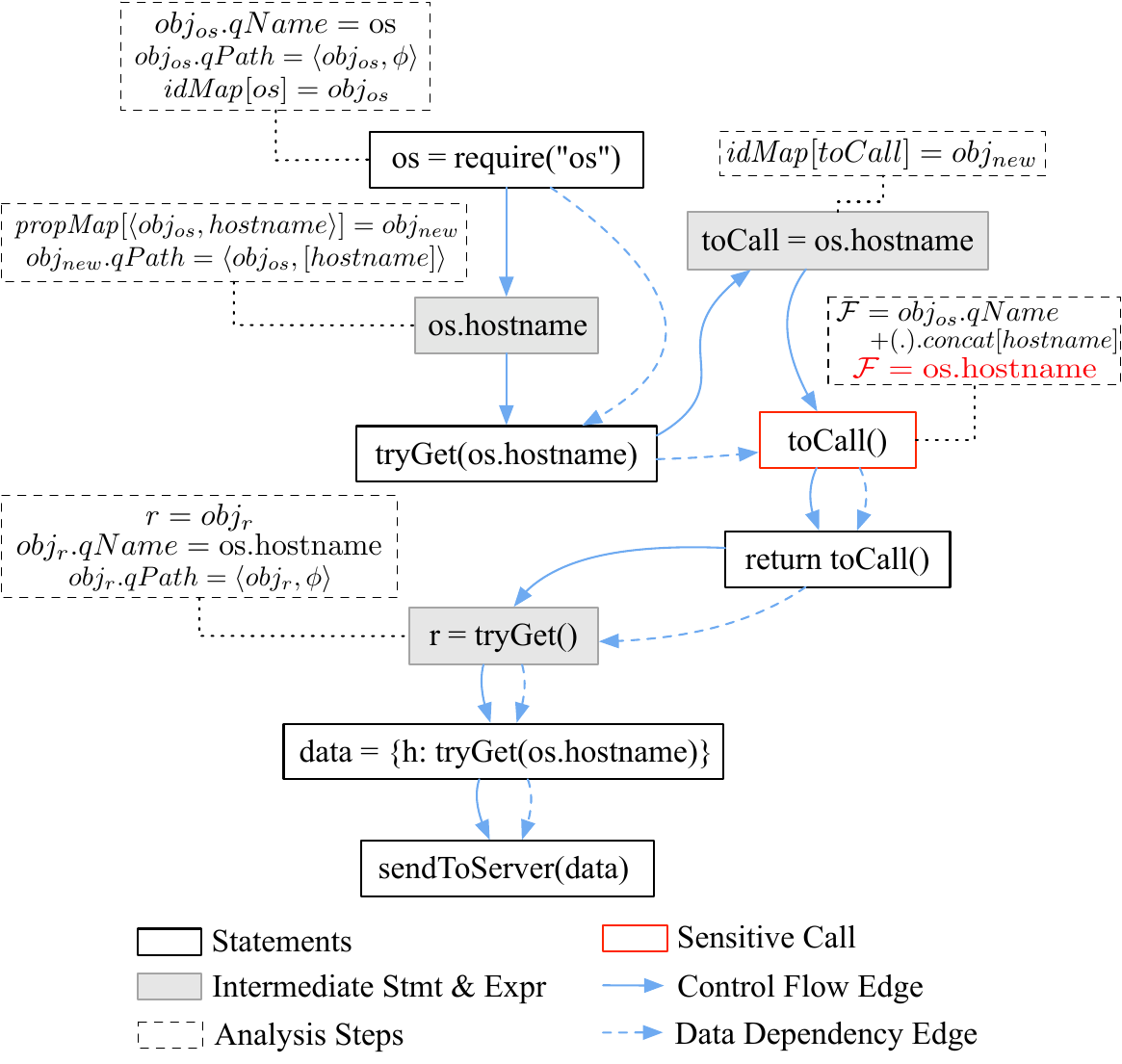}
  \vspace{-10pt}
  \caption{Behavior Graph of \textit{automation.samples\_0.1.15}}
    \label{fig:code_snippet_1}
\end{figure}

\subsubsection{An Example of BG Construction}
Fig.~\ref{fig:code_snippet_1} shows the~{partial} BG generated for the package in Fig.~\ref{fig:code_listing_1}. Specifically,~we~first analyze the assignment \texttt{os = require("os")}. Following the process for \textbf{Assignment} (Sec.~\ref{subsub:assignment}), we analyze the~RHS expression \texttt{require("os")} following the process~for~\textbf{Require} (Sec.~\ref{subsub:require}), which yields a pre-built object $obj_{os}$~representing the \texttt{os} module, with $qName = \text{``os''}$ and $qPath = \langle obj_{os}, \phi \rangle$. Then, we map the identifier to the object, $\mathit{idMap}[os] = obj_{os}$.

At the function call \texttt{tryGet(os.hostname)}, we follow the process for \textbf{Function Call} (Sec.~\ref{subsub:function_call}), resolve the callee via CG, and prepare to analyze it. Before entering~the~callee, we analyze the argument \texttt{os.hostname} following the process for \textbf{Property Access} (Sec.~\ref{subsub:property_access}). We retrieve $obj_{os}$ from $\mathit{idMap}$ for the leftmost identifier \texttt{os}, and look up the pair $\langle obj_{os}, {hostname} \rangle$ in $\mathit{propMap}$. Since the function object of \texttt{os.hostname} is not {pre-assigned}, we create a new object $obj_{new}$ with its $qPath = \langle obj_{os}, [{hostname}] \rangle$.

Then, we set the formal parameter \texttt{toCall} as an alias of $obj_{new}$, and update $\mathit{idMap}$ accordingly (i.e., $\mathit{idMap[toCall] =}$ $\mathit{obj_{new}}$). At the function call \texttt{toCall()}, we derive the full name from $obj_{new}.qPath$, which is $\mathcal{F} = obj_{os}.qName + (.).concat[{hostname}] =  \text{``os.hostname''}$. As it matches~an~entry in the sensitive API list, we mark the node as \emph{sensitive}.


\tosem{In the subsequent reasoning pipeline, the statically generated BG is converted by the code slicer (Sec.~\ref{sec:code_slicer}) into initial source-level suspicious code slices, which are then analyzed by the multi-agent reasoning framework (Sec.~\ref{sec:multi_agent_reasoning}).}

\subsection{\tosem{Code Slicer}} \label{sec:code_slicer}

\tosem{The BG preserves security-relevant contexts, but it remains an abstract representation rather than the source-code view used by human auditors. The code slicer bridges this gap in two stages. \emph{Initial graph-level slicing} starts from the statically generated BG, and produces annotated suspicious code slices. \emph{Evidence refresh} updates those slices after router-triggered augmentation; i.e., dynamic evidence from dynamic analysis (see Sec.~\ref{sec:behavior_graph_dynamic}) is read from node attributes in the augmented behavior graph and translated into inline comments, whereas documentary semantics from third-party enrichment (see Sec.~\ref{sec:third_party_enrichment}) are injected directly into the corresponding slices by node identifier without modifying the behavior graph.}

\tosem{For the initial graph-level slicing, it starts from the suspicious nodes in the BG. For a BG $G=(V,E)$, we define~the~anchor set $\mathcal{A}$ as the nodes that indicate suspicious behavior, namely (1) sensitive API nodes ($T_v=1$);~(2)~conditional~sensitive nodes ($C_v=1$); (3) statically unresolved nodes ($U_v=1$); (4) \texttt{eval} nodes ($E_v=1$); and (5) third-party API nodes ($P_v=1$).}

\tosem{For each anchor $v_a \in \mathcal{A}$, the slicer builds a graph-level slice that initially contains only the anchor, and then expands it by bidirectional slicing. Along data-dependency edges, it collects statements that define, transform, propagate, or consume values related to the anchor; i.e., backward traversal recovers the origins of receiver objects, arguments, aliases, and intermediate variables, while forward traversal captures how return values, assigned variables, and side effects are subsequently used. Along control-dependency edges, it likewise traverses bidirectionally to recover the predicates and branch structures that make the suspicious behavior feasible, such as environment checks, platform checks, and installation-time guards that decide whether a sensitive operation executes.}

\tosem{As different anchors may reach overlapping nodes, the slicer removes duplicate nodes, yielding a set of connected~subgraphs. Each subgraph represents one local suspicious behavior and serves as the basis for the following code slice.}

\subsubsection{\tosem{Code Slice Construction}}

\tosem{Given a connected subgraph, the slicer maps its nodes back to their original source statements by the location information preserved during CPG construction, and groups the nodes by file.}
\tosem{After mapping the nodes to source code, a plain list of the selected statements is often hard to read, because such statements may omit the syntax that reflects the surrounding program structure. The slicer therefore performs syntactic repair through AST, restoring enclosing block boundaries, braces, and, when a selected statement lies inside a function, the corresponding function signature. This yields coherent snippets while keeping the selected statements in their original source form.}

\tosem{For cross-function or cross-file behavior, the slicer does not concatenate distant fragments into an artificial code block. Instead, it keeps each snippet under its original file name, and records the relevant invoked functions in \texttt{call\_context}. This representation preserves both locality and cross-method structure, so that downstream agents can inspect each snippet independently while still understanding how the snippets connect.}

\subsubsection{\tosem{Inline Evidence Annotation}}

\tosem{To recover semantic information that is explicit in the BG but implicit in the raw source code, the slicer injects inline comments after key statements. Static labels (e.g., sensitive, conditional, unresolved, \texttt{eval}, and third-party call) and dynamic augmentation results (e.g., concrete arguments, return values, runtime-equivalent third-party summaries, and decoded \texttt{eval} code) are read from node attributes in the BG and translated into comments on the mapped source lines. When the router selects third-party enrichment, the enrichment agent returns module- and method-level semantics together with the target node identifier. The slicer appends these semantics directly to the corresponding statement in the annotated slice.}

\tosem{Each annotation is intentionally attached to its original statement rather than collected into a standalone metadata table, keeping the evidence next to the exact source code line that triggers it and helping the LLM better understand the code. The appended node identifier further gives the LLM a stable reference to the corresponding BG node, allowing it to locate and refer back to the key node that carries the security evidence. For example, a simplified annotated statement reads as \texttt{const hostname = os.hostname(); // Method name hostname is a sensitive API call of os.hostname. [Node ID: 30064771218]}.}

\begin{figure}[!t]
    \centering
    \includegraphics[scale=0.8]{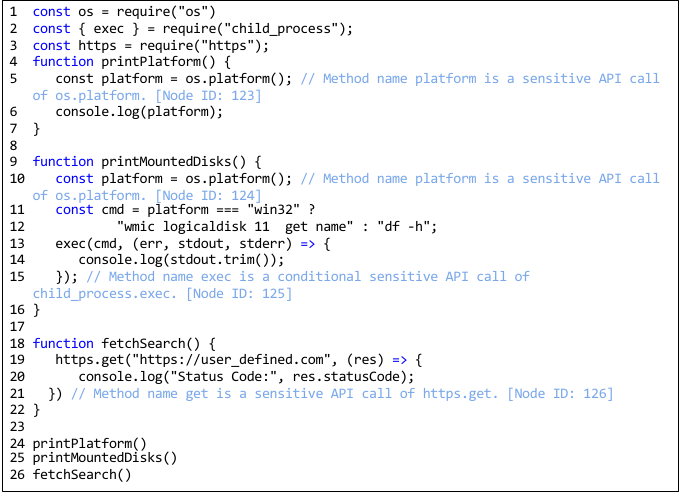}
    \vspace{-10pt}
    \caption{Code Slice Generated from the Code Snippet in Fig.~\ref{fig:code_listing_3}}
      \label{fig:code_slice_example}
  \end{figure}
  
\tosem{Fig.~\ref{fig:code_slice_example} shows the code slice generated from the code snippet in Fig.~\ref{fig:code_listing_3}. Starting from the sensitive calls \texttt{os.platform()}, \texttt{child\_process.exec()}, and \texttt{https.get()}, the slicer collects the related source statements and annotates them with the analysis facts carried by the BG, such as the resolved sensitive API name and the corresponding node identifier. In this example, every statement is reached along the data- or control-dependency edges of these anchors, so no statement is discarded and the slice coincides with the whole snippet; for larger packages, the slicer prunes statements unrelated to the anchors and retains only the security-relevant ones.}

\subsubsection{\tosem{Slice Output}}

\tosem{For each connected subgraph, the slicer outputs a \texttt{CodeSliceResult} with two fields, i.e., \texttt{code\_slice} and \texttt{call\_context}. The \texttt{code\_slice} field is organized as \texttt{\{filename: code\_snippet\}}, where each \texttt{code\_snippet} holds the source lines with their inline annotations, and \texttt{call\_context} summarizes the callees involved in the slice to help the LLM reason about cross-method behavior.}

\tosem{This output serves three downstream purposes. First, the local behavior judge agent treats each annotated slice as the unit of semantic judgement. Second, the global behavior judge agent links slice-level verdicts according to CFG order and inter-slice relations, so that malicious behavior distributed across slices can still be synthesized. Third, the localization agent reuses the slices to map the final evidence back to concrete files and statements.}


\subsection{\tosem{Multi-Agent Reasoning}} \label{sec:multi_agent_reasoning}

\tosem{After the code slicer (Sec.~\ref{sec:code_slicer}) turns the BG into a set of annotated suspicious code slices, \tool employs~coordinated LLM agents to reason over the slices and progressively accumulate evidence until a confident judgement~is~reached.}

\tosem{Rather than invoking a set of isolated agents independently, \tool organizes them around a shared and progressively enriched reasoning process. As the process proceeds, the conclusion of each agent is passed to the next stage together with its judgement rationale, key evidence, and to-be-checked nodes. In this way, each subsequent judgement is made in the context of the evidence already accumulated, instead of restarting from the raw code alone. Meanwhile, all agents operate on the same analysis carrier, i.e., the annotated code slice. Static evidence and dynamic evidence (Sec.~\ref{sec:behavior_graph_dynamic}) are injected into slices through node attributes in the BG, whereas third-party module and method semantics (Sec.~\ref{sec:third_party_enrichment}) are injected directly into slices by the code slicer without changing the BG.}

\tosem{Specifically, our multi-agent reasoning analyzes annotated slices in a closed loop. 
The local behavior judge first analyzes each slice, and produces a verified local report through self-consistency verification. Next, the global behavior~judge synthesizes the verified local reports according to the execution order induced by the CFG. If the entry-level verdict is still \texttt{undetermined}, the router selects the next evidence source, either \texttt{third\_party\_enrichment} or \texttt{dynamic\_augmentation}. The code slicer then refreshes the slice annotations using the newly collected evidence, and the local and global behavior judgements are re-invoked on the updated slices. The loop terminates when a definite verdict is reached or when no further evidence can be collected. Finally, any malicious behavior is localized in the source code files.}

\tosem{This design naturally forms a closed reasoning loop. The code slicer first produces annotated slices, and the local behavior judge agent (Sec.~\ref{subsub:local_judge}) renders a per-slice verdict by consolidating repeated judgements through self-consistency verification. The global behavior judge agent (Sec.~\ref{subsub:global_judge}) then synthesizes an entry-level verdict along the CFG order. When the entry-level verdict remains undetermined, the router agent (Sec.~\ref{subsub:router}) determines whether the current evidence gap should be addressed by third-party enrichment (Sec.~\ref{sec:third_party_enrichment}) or dynamic augmentation (Sec.~\ref{sec:behavior_graph_dynamic}). The newly obtained evidence is fed back into the corresponding slices as additional annotations, and the local and global behavior judge agents are re-invoked on the enriched slices. This loop continues until a definite malicious or benign verdict is produced, or until the available evidence is exhausted. Once a package is flagged as malicious, the localization agent (Sec.~\ref{sec:localization}) maps the malicious behavior back to concrete files and code slices.}

\subsubsection{\tosem{Prior Analysis Context}} \label{subsub:prior_context}

\tosem{A single LLM judgement over one code slice is often insufficient to draw a reliable~conclusion, because some suspicious calls may only become conclusive when combined with what earlier stages have already suspected. To this end, \tool maintains a \emph{prior analysis context} that is shared across agents and accumulated throughout the reasoning loop. The context consists of two kinds of information.}

\tosem{\textbf{Entry-trigger context.} This context records how the currently analyzed file is reached. Specifically, it distinguishes whether the file is an import-time entry that is loaded by package users, or an install-time file launched by a \texttt{preinstall}, \texttt{install}, or \texttt{postinstall} script (Sec.~\ref{sec:scripts_analyzer}). For install-time entries, the context further preserves the shell command that triggers the file and the arguments passed to it. This allows downstream agents to understand the execution condition of the current file, rather than judging it as an isolated source file. It is also necessary for detecting behaviors whose malicious intent only becomes visible through the interaction between install scripts and source code, e.g., when an install script passes a specific argument to complement a malicious intent in the launched JavaScript file.}

\tosem{\textbf{Stage-wise conclusions.} During the reasoning loop for a given entry, each stage, including local behavior judgement, global behavior judgement, the routing decision, and re-judgement, deposits its conclusion into the prior analysis context. Consequently, the next agent knows what previous stages found suspicious and why the evidence was still insufficient, instead of starting from scratch.}

\subsubsection{\tosem{Local Behavior Judge Agent}} \label{subsub:local_judge}

\tosem{The local behavior judge agent renders a verdict over a single local behavior, i.e., one annotated code slice produced by the code slicer. It determines whether the slice itself shows benign, malicious, or still inconclusive behavior, but does not combine evidence across behavior chains, which is deferred to the global behavior judge agent (Sec.~\ref{subsub:global_judge}).}

\begin{figure}[!t]
	\begin{tcolorbox}[
		title=Prompt Snippet,
		colback=white,
		colframe=black,
		fonttitle=\bfseries\footnotesize,
		boxrule=0.5pt,
		arc=2pt,
		left=1pt, right=1pt, top=1pt, bottom=1pt,
	]
	\footnotesize
	\tosem{\textbf{Task:} You are a JavaScript cybersecurity analyst. Analyze the given code slice together with the call context and the prior analysis context. Determine whether the slice exhibits malicious behavior and classify it as \texttt{benign}, \texttt{malicious}, or \texttt{undetermined}.}\\

	\tosem{\textbf{Input:} The input contains the current annotated code slice, its \texttt{<Call Context>}, and the \texttt{<Prior Analysis Context>} accumulated from earlier stages.}\\

    \tosem{\textbf{Annotations:} Each annotation is appended to a statement as an inline comment in the following format: \texttt{// Method name: <call> is a <call type> of <qualified name>. [Node ID: <N>]}. For example, \texttt{// Method name: toCall is a sensitive API call of os.hostname. [Node ID: 30064771101]}. The call type indicates whether the statement involves a sensitive API call, a conditional sensitive API call, a third-party API call, or a statically unresolved call.}\\

    \tosem{\textbf{Guidelines:} Examine the annotated slice comprehensively to identify malicious behavior. Consider trigger conditions, data and control flows, sensitive sources and sinks, network transmission, command execution, file or process manipulation, obfuscation, runtime-resolved values, and unresolved or third-party calls on security-relevant paths. Assign \texttt{malicious} only when the visible evidence is sufficient to confirm malicious intent. Representative cases include exfiltrating sensitive files or credentials to an external server, executing hardcoded reverse shells or destructive commands, downloading a remote payload and executing it, or running a native binary shipped inside the package. Assign \texttt{undetermined} when the current slice lacks enough evidence for a definitive judgement, such as when sensitive behavior depends on runtime-computed values, unresolved calls, dynamically assembled code or commands, or third-party calls without trustworthy and specific semantics.}\\

    \tosem{\textbf{Output:} Return a JSON object in the following format: \texttt{\{"judgement": "benign | malicious | undetermined", "key\_evidence": [\{"node\_id", "node\_type", "claim"\}], "reason": "<reason>", "node\_to\_be\_checked": [...]\}}. Populate \texttt{node\_to\_be\_checked} only when the judgement is \texttt{undetermined}. This field should include only the annotated nodes whose semantics may be clarified by later evidence augmentation, namely \texttt{conditional\_api}, \texttt{third\_party}, or \texttt{unresolved} nodes.}

\end{tcolorbox}

\vspace{-10pt}

\caption{\tosem{Prompt Snippet for the Local Behavior Judge Agent}}

\label{fig:local_judge_template}
\end{figure}

\tosem{The agent takes one annotated code slice as input, together with its \texttt{call\_context} and the prior analysis context. The slice contains the source statements selected by the code slicer and inline annotations that identify security-relevant calls by node identifier and call type. This compact input lets the agent focus on the local behavior while still considering how the current slice is reached and what previous stages have already concluded.}

\tosem{We instruct the LLM to first interpret the annotation format, so that it can associate each relevant statement with its node identifier and call type. We then instruct it to examine the slice comprehensively, including the annotated calls, data and control dependencies, call context, prior analysis context, and visible indicators of malicious intent, before classifying the slice as \texttt{benign}, \texttt{malicious}, or \texttt{undetermined}. We also provide in-context examples based on malicious behavior categories summarized in prior studies~\cite{ohm2020backstabber, zhang2024cerebro, huang2025ProfMal, Gao2025MALGUARD, zheng2024OSCAR}, including credential exfiltration, reverse shells or destructive commands, remote payload execution, and execution of a native binary shipped with the package. To reduce over-claiming, we instruct the LLM to follow a conservative decision policy; i.e., it reports \texttt{malicious} only when the visible slice-level evidence is sufficient, and otherwise reports \texttt{undetermined}. For an \texttt{undetermined} slice, it records only the specific annotated nodes that block a confident judgement in \texttt{node\_to\_be\_checked}. These nodes provide precise follow-up targets for later stages. Fig.~\ref{fig:local_judge_template} shows the prompt snippet used by the local behavior judge agent, and the complete prompt is available at our replication website~\cite{tool_website}.}

\tosem{Moreover, a single LLM invocation may produce different verdicts for the same code slice, and directly relying on one-shot judgement would propagate such uncertainty to downstream stages. To mitigate this source of variance, \tool wraps the local behavior judge agent with a self-consistency verification agent, following the idea of self-consistency prompting~\cite{wang2023selfconsistency}.}
\tosem{Specifically, for each local code slice, the local behavior judge agent is invoked three times independently with the same prompt, producing three candidate reports. These reports are then provided to a verification agent, which arbitrates their verdicts, reconciles their evidence, and produces a single consolidated report. The verification agent uses the same output schema as the local behavior judge agent (Fig.~\ref{fig:local_judge_template}). We instruct the verification LLM to sanitize the \texttt{node\_to\_be\_checked} field and perform a plausibility check on the candidate reports, questioning whether a suspicious or malicious verdict is sufficiently supported by the visible evidence. This self-consistency agent makes local judgements more stable, prevents unsupported or unrelated nodes from being propagated as follow-up analysis targets, and reduces false positives caused by over-interpreting incomplete evidence. The prompt snippet used by the verification agent is presented in Fig.~\ref{fig:verifier_template}, and the complete prompt is available at our replication website~\cite{tool_website}.}

\begin{figure}[!t]
	\begin{tcolorbox}[
		title=Prompt Snippet,
		colback=white,
		colframe=black,
		fonttitle=\bfseries\footnotesize,
		boxrule=0.5pt,
		arc=2pt,
		left=1pt, right=1pt, top=1pt, bottom=1pt,
	]
	\footnotesize
	\tosem{\textbf{Task:} You are a verifier that consolidates three candidate reports of the same code slice into a single revised report. Verify whether each report's judgement and key evidence are actually supported by the sliced code and its node-id annotations, and identify contradictions, over-claims, missing evidence, and false-positive risks.}\\

	\tosem{\textbf{Input:} The input contains \texttt{<Annotated Code Slice>} and \texttt{<Candidate Reports>} produced by the local behavior judge agent.}\\

	\tosem{\textbf{Guidelines:} Consolidate the three candidate reports into one verified report. Preserve a consensus only when the shared verdict is supported by the annotated slice and follows the output rules. If the majority verdict is weakly supported, internally inconsistent, or relies on assumptions beyond the visible evidence, revise it according to node-id-grounded code evidence. For major disagreements, especially conflicting \texttt{benign} and \texttt{malicious} verdicts, prioritize explicit malicious behavior patterns and false-positive checks over vote counts. When the evidence is insufficient, prefer \texttt{undetermined} over \texttt{malicious} to avoid over-reporting. Finally, ensure that the consolidated report is internally consistent: \texttt{key\_evidence} must be provided for both \texttt{malicious} and \texttt{undetermined} verdicts; \texttt{node\_to\_be\_checked} must be non-empty only when the verdict is \texttt{undetermined}, and its entries must be restricted to \texttt{conditional\_api}, \texttt{third\_party}, or \texttt{unresolved} nodes.}\\

	\tosem{\textbf{Output:} Return a single arbitrated report following the schema in Fig.~\ref{fig:local_judge_template}.}
	\end{tcolorbox}
	\vspace{-10pt}
	\caption{\tosem{Prompt Snippet for the Verification Agent}}
	\label{fig:verifier_template}
\end{figure}

\subsubsection{\tosem{Global Behavior Judge Agent}} \label{subsub:global_judge}

\tosem{After the local behavior judge agent produces stable reports for individual annotated slices through self-consistency verification, the global behavior judge agent further lifts these slice-level conclusions to an entry-level view. While the local behavior judge focuses on whether a single local behavior is malicious, benign, or still undetermined, the global behavior judge examines how multiple local behaviors under the same entry may jointly form a behavior chain. It therefore reasons over the verified local judgements, their node-grounded key evidence, and the execution-order relations induced by the CFG.}

\tosem{This step is necessary because malicious packages may deliberately distribute one attack across several local behaviors. For example, one annotated slice may only assemble a remote URL, collect an environment-dependent value, or write a temporary file, while a later slice performs an outbound request, executes a command, or loads the written artifact. Judging these slices independently may leave each of them benign or undetermined, but together they can reveal a complete malicious chain, such as credential collection followed by exfiltration, payload download followed by execution, or environment gating followed by a sensitive action.}

\tosem{The global behavior judge adopts the same conservative decision principle as the local behavior judge, but applies it at the entry level by synthesizing all local behavior reports. When any local behavior has been verified as \texttt{malicious}, this verdict is treated as sufficient evidence for a \texttt{malicious} entry-level judgement. The global behavior judge therefore preserves the local malicious conclusion by default, instead of weakening or re-interpreting it because other local behaviors appear benign. A downgrade is allowed only under a strict false-positive condition; i.e., the global context~must provide concrete contrary evidence showing that the local report misinterpreted the behavior, e.g., the allegedly sensitive operation is unreachable, guarded by a benign-only condition, or its data flow is contradicted by other slices. In such cases, the agent must explicitly explain why the local malicious evidence is invalidated. Otherwise, cross-slice reasoning is used not to revalidate the malicious verdict, but to examine whether other local behaviors extend the observed behavior into a broader malicious chain, such as a preceding credential-collection step or a subsequent exfiltration step.}

\begin{figure}[!t]
	\begin{tcolorbox}[
		title=Prompt Snippet,
		colback=white,
		colframe=black,
		fonttitle=\bfseries\footnotesize,
		boxrule=0.5pt,
		arc=2pt,
		left=1pt, right=1pt, top=1pt, bottom=1pt,
	]
	\footnotesize
	\tosem{\textbf{Task:} You are a JavaScript cybersecurity analyst performing cross-slice synthesis. Given the verified local behavior reports of one package entry and their execution-order relations, re-evaluate the overall package behavior and determine whether the entry is \texttt{benign}, \texttt{malicious}, or \texttt{undetermined}.}\\

	\tosem{\textbf{Input:} The input contains a JSON object with \texttt{<Local Behaviors>} and their execution-order relations: \texttt{<CFG Order>}. Each local behavior includes annotated code slice, call context, local judgement, key evidence, reason, and unresolved nodes reported by the local behavior judge and consolidated by the verifier. The local behaviors are organized according to the control-flow graph reachable from the program entry, and the ordering relations indicate which behaviors may execute before others.}\\

	\tosem{\textbf{Guidelines:} If any local behavior is already confirmed as \texttt{malicious}, preserve this malicious verdict for the entry by default and use cross-slice reasoning mainly to determine whether other local behaviors extend it into a broader malicious chain. Downgrade such a verdict only under a strict false-positive condition, where the global context provides concrete contrary evidence that invalidates the local malicious evidence, such as unreachable code, a benign-only guard, or a contradicted data flow. For local behaviors judged as \texttt{benign} or \texttt{undetermined}, examine whether their ordered composition forms a recognizable malicious chain, such as reading secrets followed by outbound transmission, downloading or writing a payload followed by execution, collecting credentials followed by encoding and exfiltration, environment gating followed by a sensitive action, or decoding/decrypting data that is later used by a sensitive sink. Escalate to \texttt{malicious} only when the cross-slice chain is supported by node-id-grounded evidence. If the interaction is suspicious but not conclusive, return \texttt{undetermined} and report only the unresolved, conditional, or third-party nodes whose semantics should be promoted. Do not include sensitive API, sensitive property, or already enriched third-party nodes in \texttt{node\_to\_be\_checked}. If no cross-slice malicious pattern exists and all local evidence is benign, return \texttt{benign}.}\\

	\tosem{\textbf{Output:} Return a JSON object in the following format: \texttt{\{"judgement": "benign | malicious | undetermined", "explanation": "<explanation>", "behavior\_chain\_evidence": [\{"pattern", "involved\_slices", "description"\}], "node\_to\_be\_checked": [...]\}}. Populate \texttt{node\_to\_be\_checked} only when the judgement is \texttt{undetermined}. This field should include only the annotated nodes whose semantics may be clarified by later evidence augmentation, namely \texttt{conditional\_api}, \texttt{third\_party}, or \texttt{unresolved} nodes. Use an empty array for \texttt{behavior\_chain\_evidence} when no cross-slice malicious pattern is found.}
	\end{tcolorbox}
	\vspace{-10pt}
	\caption{\tosem{Prompt Snippet for the Global Behavior Judge Agent}}
	\label{fig:global_judge_template}
\end{figure}

\tosem{When no individual local behavior is sufficient to support a \texttt{malicious} verdict, the global behavior judge reports \texttt{malicious} only if multiple local behaviors can be composed into a concrete malicious chain supported by node-identifier-grounded evidence. When such a cross-slice pattern is identified, the agent records it in \texttt{behavior\_chain\_evidence}; each entry contains the recognized \texttt{pattern}, the \texttt{involved\_slices} (identifiers of the annotated code slices), and a concise \texttt{description}. These identifiers are used downstream to map the verdict back to concrete files and statements during localization. If the composed behavior remains suspicious but does not provide decisive evidence, the agent returns \texttt{undetermined}. In this case, it promotes only the specific nodes whose unresolved semantics prevent a definitive judgement, rather than escalating the entire entry. This design preserves the node-level granularity of uncertainty from local reports, enabling the router agent (Sec.~\ref{subsub:router}) to selectively invoke third-party enrichment or dynamic augmentation in subsequent analysis. The prompt snippet used by the global behavior judge is shown in Fig.~\ref{fig:global_judge_template}, and the complete prompt is available on our replication website~\cite{tool_website}.}

\subsubsection{\tosem{Router Agent}} \label{subsub:router}

\tosem{When the global behavior judge still returns \texttt{undetermined}, the remaining uncertainty has already been localized to a set of specific nodes. The router agent is responsible for deciding which evidence source should be used next to resolve these nodes. Rather than treating all undetermined cases uniformly, it examines what information is missing for each node, and selects one of two follow-up actions, i.e., \texttt{third\_party\_enrichment}, which retrieves third-party package semantics and feeds them back into the corresponding annotated slices, or \texttt{dynamic\_augmentation}, which executes the package in a sandbox to collect dynamic evidence behind specific nodes.}

\tosem{The routing decision is driven by the type of missing information. For a \texttt{third\_party} node, the static analysis has identified the module and method, but does not know the semantics of that API. Therefore, third-party enrichment can make the node decidable by retrieving the package description, metadata, or API-level documentation when such information is specific enough to explain the invoked method. However, if the recovered semantics are ambiguous, overly generic, or come from an untrusted third-party package, the semantics cannot be determined. In this case, dynamic augmentation may be required to capture the concrete behavior triggered behind the third-party API, i.e., the underlying sensitive calls. For a \texttt{conditional\_api} node, the API semantics are already known, but the decisive argument or return value is runtime-dependent. For an \texttt{unresolved} node, the call itself cannot be resolved statically.}

\begin{figure}[!t]
	\begin{tcolorbox}[
		title=Prompt Snippet,
		colback=white,
		colframe=black,
		fonttitle=\bfseries\footnotesize,
		boxrule=0.5pt,
		arc=2pt,
		left=1pt, right=1pt, top=1pt, bottom=1pt,
	]
	\footnotesize
	\tosem{\textbf{Task:} You are the routing agent for a multi-stage JavaScript malware analysis pipeline. Previous stages have returned an \texttt{undetermined} judgement. Choose exactly one follow-up action: \texttt{third\_party\_enrichment} or \texttt{dynamic\_augmentation}.}\\

	\tosem{\textbf{Input:} The input contains \texttt{<Suspicious Nodes>} emitted by the global behavior judge, \texttt{<Annotated Slices>} containing these nodes, and \texttt{<Prior Analysis Context>}, including \texttt{<Reason>}, \texttt{<Key Evidence>}, and \texttt{<Behavior Chain Evidence>} accumulated by the local and global judgements.}\\

	\tosem{\textbf{Guidelines:} Decide which action can best supply the missing information. Prefer \texttt{third\_party\_enrichment} when most flagged nodes are specific \texttt{third\_party} calls and the missing evidence is the documented behavior of their modules or methods, because enrichment is the cheaper path and can often resolve the judgement without sandbox execution. This preference is safe because enrichment is followed by re-judgement, and dynamic augmentation is still invoked if the recovered third-party evidence is unavailable, too generic, untrusted, or still insufficient for a definitive verdict. Choose \texttt{dynamic\_augmentation} directly when the uncertainty is dominated by \texttt{conditional\_api} or \texttt{unresolved} nodes, where the missing evidence is the concrete runtime argument, return value, or callee identity. Also choose \texttt{dynamic\_augmentation} when the annotated slice shows that the decisive evidence is a realized command string, file path, URL, decoded payload, or transitive runtime behavior that documentation cannot determine. Ground the decision in the flagged nodes and prior analysis, not only in node counts.}\\

	\tosem{\textbf{Output:} Return a JSON object in the following format: \texttt{\{"next\_action": "third\_party\_enrichment | dynamic\_augmentation", "reason": "<reason>"\}}. The \texttt{next\_action} must be exactly one of these two values. The \texttt{reason} must explain what information is missing and why the selected action can supply it.}
	\end{tcolorbox}
	\vspace{-10pt}
	\caption{\tosem{Prompt Snippet for the Router Agent}}
	\label{fig:router_template}
\end{figure}

\tosem{Since third-party enrichment is substantially cheaper than dynamic execution, the router is biased toward third-party enrichment whenever the remaining uncertainty is mainly caused by specific third-party calls whose documented behavior is likely to clarify the judgement. This preference is safe because enrichment is not treated as a terminal shortcut. If the recovered documentation is unavailable, or rejected as untrustworthy, and the entry remains \texttt{undetermined} after re-judgement, \tool falls through to dynamic augmentation. In contrast, the router selects dynamic augmentation directly when the suspicious nodes are dominated by \texttt{conditional\_api} or \texttt{unresolved} calls.}

\tosem{The router takes as input the suspicious nodes emitted by the global behavior judge, the annotated slices that contain these nodes, and the prior analysis context, including the reasons, key evidence, and behavior-chain evidence accumulated by the local and global judgements. It returns a single next action, \texttt{third\_party\_enrichment} or \texttt{dynamic\_augmentation}, together with a concise justification grounded in the flagged nodes. After the selected action promotes the evidence, the updated annotations are fed back into the corresponding slices, and the local and global behavior judge agents are re-invoked. If enrichment fails to resolve the uncertainty because the third-party evidence is insufficient or untrusted, dynamic augmentation is invoked next; otherwise, the loop stops once a definite judgement is reached or no further evidence can be collected. The prompt snippet used by the router agent is presented in Fig.~\ref{fig:router_template}, while the complete prompt is available at our replication website~\cite{tool_website}.}


\subsection{\tosem{Third-Party Enrichment Agent}} \label{sec:third_party_enrichment}

\tosem{Using third-party libraries is a common practice in modern software packages, where package code often delegates substantial functionality to external dependencies rather than implementing it locally. However, from the local package code alone, a call site usually only reveals that an external method is invoked, but not the semantics of that method.~As~shown in Sec.~\ref{sec:behavior_graph_static}, the static generator already resolves such call sites to a third-party node ($P_v=1$) and records the module name, declared version, and method path (e.g., module \texttt{axios} and method \texttt{post}). The third-party enrichment agent is invoked on demand when the router agent (Sec.~\ref{subsub:router}) selects the \texttt{third\_party\_enrichment} action. It takes each flagged third-party node together with its resolved \emph{(module, method, node\_identifier)} tuple, retrieves external semantic evidence, and returns the result to the code slicer (Sec.~\ref{sec:code_slicer}), which injects it into the corresponding annotated slice.}

\tosem{Unlike prior work such as \textsc{SpiderScan}~\cite{huang2024spiderscan} that locates an exposed API and parses its function body to characterize behavior, the goal of this agent is to recover the semantics of an external API as a developer would understand it, not its implementation details. Additionally, for most NPM packages, the public-facing API is wrapped through multiple layers (entry-point wrappers or adapters), so mapping a method name to a concrete function body may introduce noise and can be unreliable. We instead rely on lightweight, developer-facing documentary evidence, namely registry metadata, public documentation, and declaration files, which is closer to call-site semantics than the implementation details.}

\subsubsection{\tosem{Trustworthiness Assessment}}

\tosem{Before retrieving third-party semantics, the agent first performs trustworthiness assessment because package-provided documentation and metadata are not inherently reliable. In particular, an attacker may control the documentation and metadata of a package they publish. Blindly trusting such self-descriptions would allow a malicious package to present its own malicious behavior as a benign utility. The trustworthiness assessment therefore serves as a safeguard against semantic poisoning, ensuring that only self-descriptions supported by independent ecosystem signals are retrieved and used as external semantic evidence.}

\begin{figure}[!t]
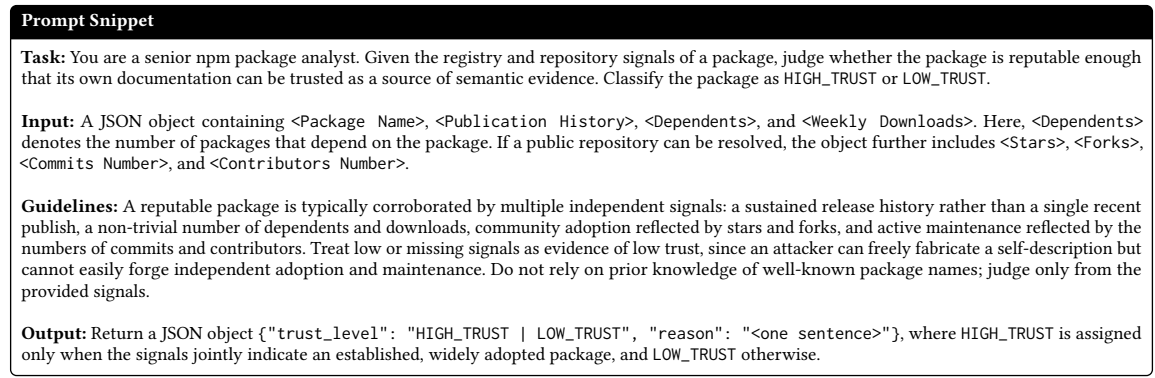

	\begin{tcolorbox}[  
		title=Prompt Snippet,
		colback=white,
		colframe=black,
		fonttitle=\bfseries\footnotesize,
		boxrule=0.5pt,
		arc=2pt,
		left=1pt, right=1pt, top=1pt, bottom=1pt,
	]
	\footnotesize
	\tosem{\textbf{Task:} You are a senior npm package analyst. Given the registry and repository signals of a package, judge whether the package is reputable enough that its own documentation can be trusted as a source of semantic evidence. Classify the package as \texttt{HIGH\_TRUST} or \texttt{LOW\_TRUST}.}\\

	\tosem{\textbf{Input:} A JSON object containing \texttt{<Package Name>}, \texttt{<Publication History>}, \texttt{<Dependents>}, and \texttt{<Weekly Downloads>}. Here, \texttt{<Dependents>} denotes the number of packages that depend on the package. If a public repository can be resolved, the object further includes \texttt{<Stars>}, \texttt{<Forks>}, \texttt{<Commits Number>}, and \texttt{<Contributors Number>}.}\\

	\tosem{\textbf{Guidelines:} A reputable package is typically corroborated by multiple independent signals: a sustained release history rather than a single recent publish, a non-trivial number of dependents and downloads, community adoption reflected by stars and forks, and active maintenance reflected by the numbers of commits and contributors. Treat low or missing signals as evidence of low trust, since an attacker can freely fabricate a self-description but cannot easily forge independent adoption and maintenance. Do not rely on prior knowledge of well-known package names; judge only from the provided signals.}\\

	\tosem{\textbf{Output:} Return a JSON object \texttt{\{"trust\_level": "HIGH\_TRUST | LOW\_TRUST", "reason": "<one sentence>"\}}, where \texttt{HIGH\_TRUST} is assigned only when the signals jointly indicate an established, widely adopted package, and \texttt{LOW\_TRUST} otherwise.}
	\end{tcolorbox}
	\vspace{-10pt}
	\caption{\tosem{Prompt Snippet for Third-Party Trustworthiness Assessment}}
	\label{fig:trustworthiness_template}
\end{figure}

\tosem{Concretely, the agent conducts a module-level trustworthiness assessment using NPM registry signals, including publication history, dependent information, and download counts. When the resolved repository points to a public code host such as GitHub, the agent further collects repository signals, including the numbers of stars, forks, commits, and contributors, as auxiliary popularity and maintenance evidence. Rather than relying on hard-coded thresholds over these heterogeneous signals, we instruct the LLM to judge whether the module is trustworthy enough for its self-description to be used as semantic evidence. The prompt snippet used for trustworthiness assessment is presented in Fig.~\ref{fig:trustworthiness_template}, while the complete prompt is available at our replication website~\cite{tool_website}. If a module fails the assessment, the agent returns an untrusted-third-party marker together with the node identifier, and the code slicer records this marker in the corresponding slice annotation; otherwise, the agent proceeds to the subsequent semantic inference.}

\subsubsection{\tosem{Documentation Candidate Discovery}}

\tosem{For a trusted module, the agent collects documentary evidence from the \texttt{package.json} of the corresponding dependency, organized into three groups. (1) \emph{Module-level fields}: \texttt{description} and \texttt{keywords} summarize the overall purpose of the module, allowing the LLM to place it into a coarse category such as HTTP client, utility library, parser, or crypto library. (2) \emph{Documentation-entry fields}: \texttt{homepage} and \texttt{repository} are used to discover documentation pages, from which the agent fetches and parses the textual content; these pages may point to a GitHub repository, allowing the agent to further locate \texttt{README} files. (3) \emph{Interface-level fields}: \texttt{types}, \texttt{typings} and \texttt{exports.types} that point to the TypeScript declaration files, which carry structured interface information (e.g., parameter and return types and the associated JSDoc comments) and thus provide the most direct evidence for method-level semantics. Since these declaration files can be lengthy, the agent does not feed the whole file to the LLM. Instead, it parses each file into an AST and, guided by the known method name, extracts only the declaration blocks that define or document that method as the method-level evidence, including its enclosing function declaration, interface/class method, object-property method, and the attached JSDoc comment.}

\subsubsection{\tosem{Evidence-Constrained Semantic Inference}}

\tosem{The collected evidence is assembled and handed to the LLM, which distinguishes the role of each source. Specifically, \texttt{description} and \texttt{keywords} determine the overall function of the module; \texttt{homepage}, \texttt{repository}, and the fetched \texttt{README} text provide documentation context and usage scenarios; and the extracted declaration evidence determines the parameters, return value, and interface semantics of the specific method. Based on these sources, the agent produces two layers of semantics, i.e., a \emph{module-level} description of what the module does overall, and a \emph{method-level} description of what the invoked method does.}

\tosem{Inference is strictly evidence-constrained to prevent the LLM from substituting prior knowledge for missing facts. The agent is instructed to (1) output only semantics supported by the evidence package; (2) refrain from completing missing semantics from pre-training knowledge; (3) emit only the module-level semantics and mark the method-level semantics as \texttt{unknown} when the evidence determines the module but is insufficient for the method; and (4) mark both layers as \texttt{unknown} when neither can be reliably grounded. This conservative policy avoids unreliable interpretations produced from prior knowledge. The prompt snippet used for evidence-constrained semantic inference is presented in Fig.~\ref{fig:enrichment_template}, while the complete prompt is available at our replication website~\cite{tool_website}.}

\begin{figure}[!h]
	\begin{tcolorbox}[
		title=Prompt Snippet,
		colback=white,
		colframe=black,
		fonttitle=\bfseries\footnotesize,
		boxrule=0.5pt,
		arc=2pt,
		left=1pt, right=1pt, top=1pt, bottom=1pt,
	]
	\footnotesize
	\tosem{\textbf{Task:} You are a senior Node.js package analyst. Given the metadata of a high-trust package and a target method name, produce a concise high-level description of the module's overall functionality and of what the invoked method does.}\\

	\tosem{\textbf{Input:} A JSON object with the target \texttt{<Method Name>} and three groups of evidence: \emph{module-level fields} (\texttt{<Description>}, \texttt{<Keywords>}), \emph{documentation-entry fields} (the text fetched from \texttt{<Homepage>} and \texttt{<Repository>}), and \emph{interface-level fields} (the \texttt{<Declaration Evidence>} extracted from the declaration files).}\\

	\tosem{\textbf{Evidence Rules:} Base the answer only on the provided fields and do not rely on outside knowledge about well-known packages. For the module-level semantics, prioritize evidence as \emph{documentation-entry} $>$ \emph{module-level} $>$ \emph{interface-level}. For the method-level semantics, combine the \emph{interface-level} declaration evidence with the relevant API references, signatures, and usage examples located in the \emph{documentation-entry} content, falling back to the \emph{module-level} fields only to disambiguate. Describe security-relevant effects (network, filesystem, process, system information) when clearly indicated. If the evidence is too vague to support a confident description, output \texttt{unknown}.}\\

	\tosem{\textbf{Output:} Return a JSON object \texttt{\{"module\_functionality": "<description>", "api\_behavior": "<description>"\}}, where \texttt{module\_functionality} is the module-level semantics and \texttt{api\_behavior} is the method-level semantics, each a single sentence, and either value is set to \texttt{unknown} when its evidence is insufficient.}
	\end{tcolorbox}
	\vspace{-10pt}
	\caption{\tosem{Prompt Snippet for Evidence-Constrained Semantic Inference}}
	\label{fig:enrichment_template}
\end{figure}

\tosem{Finally, the agent returns the two-layer description together, and the code slicer locates the call statement by node identifier and appends the semantics as an inline annotation on the corresponding slice.}


\subsection{Behavior Graph Dynamic Augmentor} \label{sec:behavior_graph_dynamic}

\tosem{The dynamic augmentor is invoked on demand by the router agent (Sec.~\ref{subsub:router}) as the \texttt{dynamic\_augmentation} action, and is triggered only for the specific suspicious nodes routed to it, namely \emph{conditional} nodes, \emph{unresolved} nodes, \emph{eval} nodes, and \emph{third-party} nodes whose semantics failed to be resolved by third-party enrichment.}~In~line with prior work~\cite{huang2024donapi}, we focus on the installation~and~import phases, which are known hotspots for malicious code injection \cite{huang2024donapi, ohm2020backstabber}, and monitor at the Node.js API level during execution, excluding low-level system calls to reduce noise. \tosem{During a single execution, the augmentor records runtime facts for the routed nodes, and then feeds the collected evidence back into the BG to complete these nodes.}

When invoked, the augmentor executes the package from its designated entry file and instruments the JavaScript runtime to collect evidence for the routed nodes. During such an instrumented execution, we generate a dynamic call graph (DCG),~a sensitive call log (SCL), and an Eval log (EL). The DCG~contains call entries $\langle caller, callee \rangle$, where $caller$~is~the~call~statement and $callee$ is the callee’s definition location. The~SCL contains entries $\langle caller, \langle f, a, r \rangle \rangle$, where $f$ is the full name~of the called sensitive API, $a$ is the arguments, and $r$ is the return value. The EL contains entries $\langle caller, code \rangle$, where $code$ is the actual dynamically generated code passed to \texttt{eval}. In these three sets, $caller$ includes precise source location information, enabling direct mapping to nodes in the BG and {DDG}$^{+}$. 

We record each function call including the function~\texttt{eval}, and thereby construct the DCG and EL. Besides, we modify~the Node.js framework's builtin module layer by instrumenting each sensitive function to log the $caller$, full name ($f$),~arguments ($a$), and return value ($r$); and for asynchronous functions (e.g., \texttt{fs.readFile}), we instrument the callback handlers to capture returned data. In this way, we construct the SCL.


\tosem{Given the routed nodes, the augmentor completes them from the captured runtime facts as follows.} For each \emph{conditional} node $v$ ($C_v = 1$), we find a match~in~the SCL~based on the $caller$. If found, we record the concrete arguments and return value from the SCL as runtime context for the node; otherwise, the node is left unchanged.

For each \emph{unresolved} node $v$ ($U_v = 1$), we look for its mappings to DCG's $caller$. If none exists, it indicates that no \emph{unresolved} node corresponds to an unresolved user-defined function. Then, we search each \emph{unresolved} node in the BG for a match in the SCL, seeking unresolved sensitive API calls. Each matched node $v$ is set to \emph{sensitive} ($T_v = 1$), and those unmatched nodes are left unchanged.

If mappings to the DCG’s $caller$ exist, it indicates~the~presence of unresolved user-defined function calls. In this case, each \emph{unresolved} node in the BG that does not match an entry in the SCL is mapped to its corresponding $caller$ in the~DCG. If multiple \emph{unresolved} nodes are mapped, we identify the control-dominant node and regenerate the BG from that point, ensuring flow-sensitive and accurate analysis, especially when new behaviors appear in previously unvisited functions.

\tosem{For each \emph{third-party} node $v$ ($P_v = 1$) routed to dynamic augmentation, stepping into the library to resolve its behavior would be prohibitively expensive. Unlike documentation-based third-party enrichment (Sec.~\ref{sec:third_party_enrichment}), which injects registry-derived semantics directly into slices, dynamic augmentation operates on the behavior graph. Instead, we reconstruct the sequence of sensitive APIs that the third-party call actually triggers at runtime. From the SCL and DCG, we collect the sensitive calls that occur within the execution of the third-party call site, preserving their semantics, arguments, return values, and execution order. This runtime-equivalent sensitive API sequence is passed to the LLM, which summarizes it into a one-sentence description of the underlying behavior, with emphasis on security-relevant effects. The summary is attached as a node attribute on the third-party node in the augmented BG, so that a library call is replaced by its runtime-equivalent behavior in the refreshed slice. The complete prompt of the API-sequence interpreter is available at our replication website~\cite{tool_website}.}

A similar process applies to dynamically generated code via \texttt{eval}. If \emph{eval} nodes in the BG can be mapped to EL entries, we inject the corresponding code for each mapped node. \tosem{The decoded code passed to \texttt{eval} is additionally recorded as an annotation on the \texttt{eval} node, so that the dynamically executed code becomes directly visible in the subsequent slice.}

Before regenerating the BG for the routed nodes, we merge the DCG with the CG to form the merged call graph (MCG), combining dynamic precision with static completeness for more accurate analysis.

During this regeneration step, we enhance the process in two ways. First, we mark nodes as \emph{sensitive} if they appear in the SCL. This includes newly~discovered API calls that could not be resolved statically but are found in the traversal of previously unvisited functions in the DCG, ensuring no sensitive calls are missed. Second, we replace the CG with the MCG to add inter-procedural edges, improving~accuracy. \tosem{Because regeneration can traverse previously unvisited functions, dynamic augmentation may reveal new execution paths and callees; the BG structure therefore changes, and the newly discovered nodes are completed with the same dynamic evidence as above.}

\tosem{Finally, the augmentor returns an updated BG in which the routed nodes are completed with dynamic evidence, i.e., concrete arguments and return values for conditional nodes, resolved sensitive APIs for unresolved nodes, runtime-equivalent behavior summaries for third-party nodes, and decoded code for \texttt{eval} nodes, together with any newly discovered execution paths. These results are carried as node attributes on the augmented BG. The code slicer (Sec.~\ref{sec:code_slicer}) reads the updated attributes and refreshes the inline annotations on the affected slices, after which the local and global behavior judge agents can re-judge the enriched evidence.}

\begin{figure}[!t]
	\centering
	\includegraphics[scale=0.369]{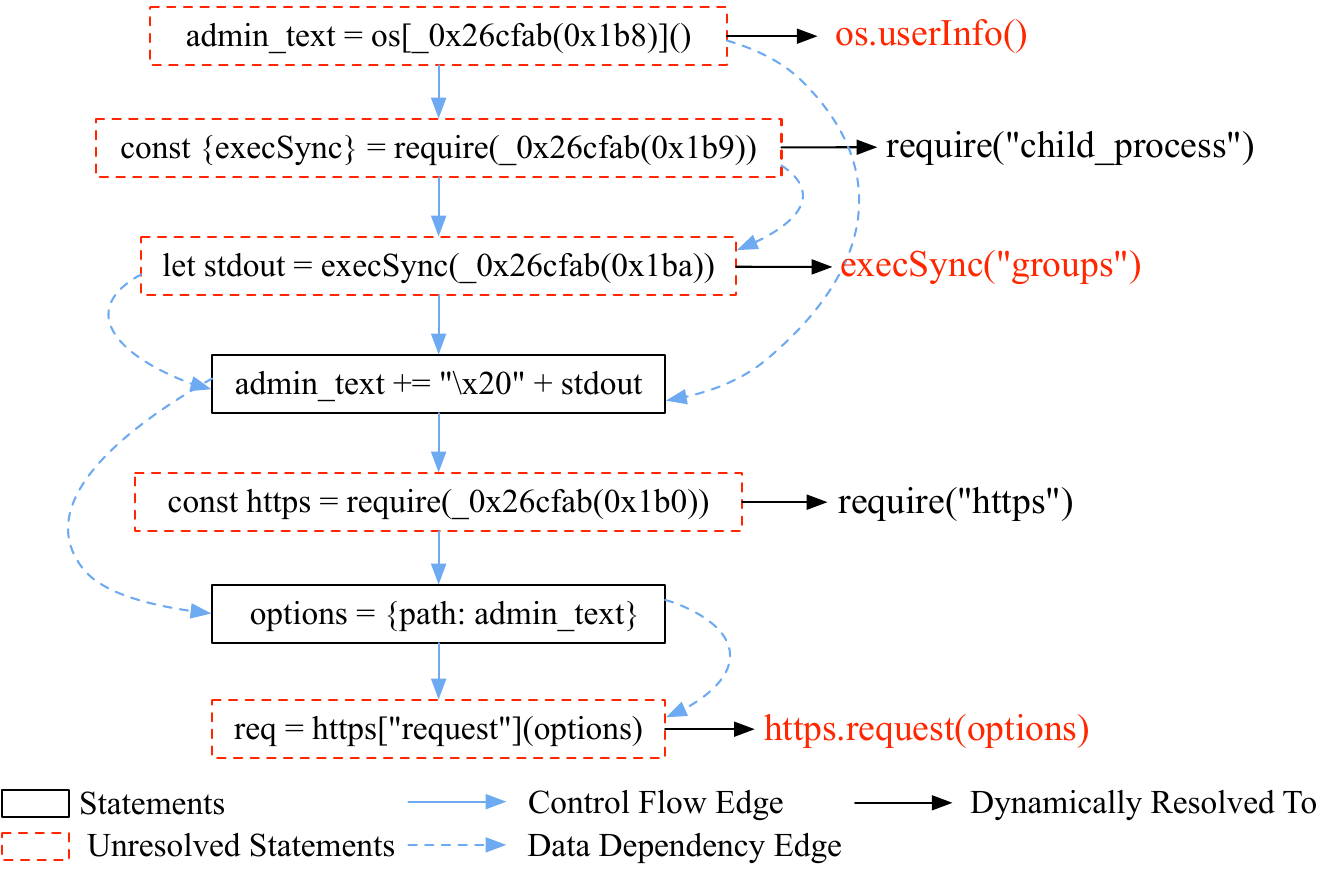}
	\vspace{-10pt}
 	\caption{Router-Triggered Augmented BG of \textit{bitsoex\_react-design-system\_14.1.4}}
    \label{fig:code_snippet_2}
\end{figure}

Fig.~\ref{fig:code_snippet_2} illustrates the augmented BG produced after the router selects dynamic augmentation for the package in Fig.~\ref{fig:code_listing_2}. Statements outlined with dashed red~boxes indicate statically unresolved calls. For the routed nodes, the dynamic augmentor resolves these calls to their runtime counterparts, as indicated by the black arrows with sensitive API calls highlighted in red.



\begin{figure}[!t]
	\begin{tcolorbox}[
		title=Prompt Snippet,
		colback=white,
		colframe=black,
		fonttitle=\bfseries\footnotesize,
		boxrule=0.5pt,
		arc=2pt,
		left=1pt, right=1pt, top=1pt, bottom=1pt,
	]
	\footnotesize
	\tosem{\textbf{Task:} You are a JavaScript cybersecurity analyst performing malicious code localization for a package that the pipeline has already judged \texttt{malicious}. Extract the source snippets that constitute the malicious behavior, copied verbatim from the provided code slices. Do not re-judge whether the package is malicious.}\\

	\tosem{\textbf{Input:} The input contains the synthesized malicious verdict, including \texttt{<Verdict>}, \texttt{<Key Evidence>}, and \texttt{<Behavior Chain Evidence>}, together with \texttt{<Annotated Slices>} implicated by that verdict.}\\

	\tosem{\textbf{Guidelines:} Starting from the statements carrying the cited key-evidence nodes, locate the code that directly implements each malicious behavior and expand outward to the complete behavioral unit (payload construction, encoding, guarding condition, and the sensitive call itself). Copy the corresponding source verbatim from the slice, stripping the injected analysis annotations so the snippet matches the original source. Restrict file names to files that appear in the provided slices, emit a separate location for each non-contiguous block, and ensure every malicious behavior cited in the verdict is covered. Do not paraphrase, abridge, or re-judge.}\\

	\tosem{\textbf{Output:} Return a JSON object in the following format: \texttt{\{"package", "entry", "summary": "<overall malicious behavior>", "locations": [\{"file", "code"\}]\}}. The \texttt{summary} field describes the overall malicious behavior of the entry. Each \texttt{locations} entry gives the file name and the verbatim malicious snippet.}
	\end{tcolorbox}
	\vspace{-10pt}
	\caption{\tosem{Prompt Snippet for the Localization Agent}}
	\label{fig:localization_template}
\end{figure}

\subsection{\tosem{Localization Agent}} \label{sec:localization}

\tosem{After the global behavior judge agent determines that an entry is \texttt{malicious}, \tool invokes the localization agent to produce human-readable evidence. The goal of this agent is not to re-judge the entry, but to map the final malicious verdict back to the concrete source code that implements the behavior. We provide the agent with the final verdict, the node-grounded key evidence, the behavior-chain evidence, and the annotated code slices related to these evidence nodes. Instead of scanning the whole package, the agent only examines the slices that were used by previous agents to support the malicious judgement. This keeps localization aligned with the reasoning process and avoids reporting code that was not involved in the final decision.}

\tosem{The agent starts from the statements that contain the cited key-evidence nodes and expands to the surrounding code needed to understand the behavior. Such code may include payload construction, encoding, guarding conditions, and the sensitive API call. For each localized behavior, the agent copies the source code from the slice and removes the inline analysis annotations added during slicing and enrichment, so that the reported snippet matches the original source file. We also instruct the agent to use only file names that appear in the analyzed slices, and cover every malicious behavior mentioned in the final verdict. The output contains a summary of the malicious behavior and a list of localization entries, each including the file name and the verbatim code snippet. The prompt snippet used by the localization agent is presented in Fig.~\ref{fig:localization_template}, while the complete prompt is available at our replication website~\cite{tool_website}.}


\section{Evaluation}

We have implemented \tool in 15K lines of Python~code. The CPG is generated by Joern~\cite{Joern}, and the CG is produced~by Jelly~\cite{Jelly}. For dynamic analysis, we~adopt NodeProf~\cite{NodeProf},~and containerize the dynamic execution environment via Docker~\cite{Docker}. \tosem{All LLM-related tasks use DeepSeek-V4-Flash~\cite{deepseek-v4-Flash} as the default backbone. 
}

We design five research questions to evaluate \tool.

\begin{itemize}[leftmargin=*]
  \item \textbf{RQ1 Effectiveness Evaluation:} How effective is \tool compared to state-of-the-art detectors?
  \item \tosem{\textbf{RQ2 LLM Backbone Selection:} How do candidate LLM backbones trade off detection effectiveness and monetary cost under the same \tool pipeline?}
  \item \textbf{RQ3 Ablation Study:} What is the contribution of each key component to \tool's effectiveness? 
  \item \tosem{\textbf{RQ4 Interpretability Evaluation:} How accurately does \tool localize malicious behavior to concrete code snippets and describe its malicious type?}
  \item \textbf{RQ5 Usefulness Evaluation:} How useful is \tool in the real-world detection scenario?
\end{itemize}

\subsection{Evaluation Setup}

\textbf{NPM Dataset.} The NPM dataset consists of both malicious and benign packages. The malicious portion is derived from publicly available datasets, including Malware Bench~\cite{zahan2024malwarebench}, Backstabber’s~Knife~Collection~\cite{ohm2020backstabber}, \textsc{MalOSS}~\cite{duan2020maloss}, and \textsc{OSCAR}~\cite{zheng2024OSCAR}. We initially collected a total of \todo{13,467} malicious packages, and deduplicated them first by package name. To further~eliminate redundancy caused by code reuse across packages, \modified{we performed file-level deduplication via perfect code matching, as many malicious packages are identical except for their version numbers}, resulting in \todo{1,658} packages. We then manually filtered out packages based on the following criteria:
{(1)} packages that contained no actual malicious code (e.g., those created purely as security placeholders or proof-of-concepts);
{(2)} obfuscated packages that fail to execute in our Linux sandbox;
{(3)} packages whose malicious behavior could not be triggered because they were not embedded in installation or import phases. 
Regarding (2), we exclude obfuscated packages that fail to execute in our sandbox, because their reported malicious behaviors cannot be reproduced or verified under our evaluation setting.
After this curation process, we retained \todo{1,090} malicious packages.

The benign dataset is drawn from three sources: {(1)} the top 5,000 NPM packages by download count, following previous work~\cite{zhang2024cerebro,huang2024spiderscan}; {(2)} benign packages from Malware Bench~\cite{zahan2024malwarebench}; and {(3)} benign packages from \textsc{OSCAR}~\cite{zheng2024OSCAR}. We applied the behavior-extraction stage of \tool during both package installation and import, and retained only packages that exposed at least one suspicious anchor. Packages with empty behavior graphs across these phases were removed: they provide no behavior slices to \tool and no graphs to graph-based baselines, and thus would be trivially classified as benign without exercising either the multi-agent reasoning pipeline or the graph classifier. After filtering, we retained \todo{3,000} benign packages.


\textbf{Baseline Selection.}
\tosem{To compare the effectiveness of \tool, we selected one rule-based detector \textsc{GuardDog}~\cite{GuardDog}, three learning-based detectors \textsc{Cerebro}~\cite{zhang2024cerebro}, \textsc{Malpacdetector}~\cite{wang2025Malpacdetector}, and \textsc{ProfMal}~\cite{huang2025ProfMal}, and one LLM-based detector, \textsc{SocketAI}~\cite{zahan2024detectionByLLM}. These baselines are representative state-of-the-art detectors in different technical categories. We restrict our selection to detectors that are either open-source or straightforward to implement from their published descriptions. \textsc{ProfMal} is included to quantify the gain from replacing its GNN classifier with our multi-agent reasoning framework. For \textsc{SocketAI}, a score above 0.5~in~its~report~indicates a malicious result. In \textsc{GuardDog}, detection is flagged as malicious if any potential malicious alarm is raised. \textsc{SocketAI} originally relies on GPT-3 from the GPT family, in our experiments, we replace its LLM backends with GPT-5.4-mini~\cite{gpt-5-4-mini}. We do not include \textsc{EMPHunter}~\cite{liang2025ClusteringInstallationScripts} in the main comparison, because its design goal and evaluation protocol differ from ours. \textsc{EMPHunter} is a clustering-and-ranking approach for installation scripts; i.e., it identifies outlier scripts from newly uploaded packages, ranks them as suspicious candidates, and relies on analysts to inspect the top-$k$ results. Accordingly, its original evaluation reports top-$k$ auditing results and mAP, which better reflects a real-world malware detection or analyst-assisted screening scenario. In contrast, our evaluation follows a package-level classification setting, where each baseline is expected to produce a malicious or benign verdict for every package so that precision, recall, and F1-score can be computed under the same protocol. Directly converting \textsc{EMPHunter}'s ranked candidates into binary predictions would require choosing an additional top-$k$ threshold, making the results sensitive to this arbitrary choice and inconsistent with its intended usage. Moreover, \textsc{EMPHunter} focuses on installation-time behaviors, whereas our evaluation covers both installation-time and import-time behaviors. Therefore, we compare \textsc{EMPHunter} only in real-world detection (Sec.~\ref{sec:usefulness-evaluation}), rather than include it in the effectiveness evaluation.}

\subsection{Effectiveness Evaluation (RQ1)}\label{sec:effectiveness-evaluation}

\textbf{RQ1 Setup.} For learning-based detectors, we conducted~10-fold cross-validation and report the average precision, recall, and F1-score across the ten folds. Rule-based and LLM-based detectors, including \textsc{GuardDog}, \textsc{SocketAI}, and \tool, were evaluated once on the entire dataset under the same package-level classification protocol. Effectiveness was measured by precision, recall, and F1-score. \tosem{Since \textsc{SocketAI} and \tool are LLM-based detectors, and \textsc{ProfMal} also incorporates LLM-based components in its pipeline, we additionally report their total cost on the evaluation dataset; the remaining baselines incur no API expense and are denoted as not applicable (``--''). We also report the average end-to-end analysis time per package for all detectors. For \textsc{SocketAI}, based on ChatGPT's rate limits~\cite{chgatgpt_rate_limit} and the average file counts of our dataset, we set the number of concurrently analyzed files per package to 10. All detectors ran on an Ubuntu server equipped with an Intel Xeon Silver 4316 CPU, 256 GB of RAM, and an RTX 3090 GPU. All results are summarized in Table~\ref{tab: tools_effectiveness}.}

\begin{table}[!t]
  \centering
  \small
  \caption{Results of Effectiveness Evaluation}
  \vspace{-10pt}
  \label{tab: tools_effectiveness}
    \tosem{%
    \begin{tabular}{
      >{\centering\arraybackslash}m{1.85cm}
      >{\centering\arraybackslash}m{1.25cm}
      >{\centering\arraybackslash}m{1.25cm}
      >{\centering\arraybackslash}m{1.25cm}
      >{\centering\arraybackslash}m{1.35cm}
      >{\centering\arraybackslash}m{1.25cm}
    }
      \toprule
      \textbf{Tool} & \textbf{Precision} & \textbf{Recall} & \textbf{F1-Score} & \textbf{Time (s)} & \textbf{Cost}  \\
      \midrule
      \textsc{GuardDog}           &  49.2\%       & 93.2\%      & 64.3\%      & 151.88           & --        \\ 
      \textsc{Cerebro}            &  83.5\%       & 80.2\%      & 81.8\%      & 25.64            & --        \\ 
      \textsc{Malpacdetector}     &  93.6\%       & 88.6\%      & 91.0\%      & \textbf{1.44}    & --        \\ 
      \textsc{SocketAI}           &  95.0\%       & 94.5\%      & 94.8\%      & 176.91           & \$562.48  \\ 
      \textsc{ProfMal}            &  91.4\%       & 90.3\%      & 90.9\%      & 192.71           & \$19.43   \\ 
      \tool                       &  \textbf{98.0}\%       & \textbf{98.2}\%      & \textbf{98.1}\%      & 394.37      & \textbf{\$67.45}      \\ 
      \bottomrule
    \end{tabular}%
    }%
  \end{table}

\tosem{\textbf{Overall Effectiveness Results.}}
\tosem{Table~\ref{tab: tools_effectiveness} reports the overall effectiveness results. \tool achieves the best performance, with an F1-score of \todo{98.1\%}, outperforming the state-of-the-art detectors by \todo{3.5\%} to \todo{52.6\%}. \textsc{SocketAI} is the strongest baseline, achieving an F1-score of \todo{94.8\%}, and only slightly trails \tool in both precision and recall. \textsc{MalPacDetector} also obtains strong precision and F1-score, demonstrating its effectiveness in identifying malicious packages with relatively few false positives. In contrast, \textsc{GuardDog} achieves high recall but substantially lower precision, suggesting that its rule-based alarms are sensitive to malicious behaviors but tend to produce many false positives. \textsc{Cerebro} shows moderate effectiveness among the learning-based baselines.}

\tosem{\textbf{False Positive and False Negative Analysis.}}
\tosem{\tool's false negatives mainly come from incomplete~behavior extraction or unavailable dynamic evidence. First, some malicious packages embed malicious logic in large bundled files or generated code that exceeds the parsing capability of Joern, so the corresponding behavior nodes cannot be extracted. Second, several packages download code from the network and execute it through \texttt{eval}, when dynamic execution cannot be triggered, the downloaded code remains unknown and the suspicious behavior is conservatively downgraded. Third, some packages use webpack-style bundling or complex trigger conditions, causing the malicious logic to be missed by our extraction pipeline. Besides, \tool's false positives mainly come from highly suspicious behaviors whose benign intent cannot be verified from source-level evidence. Some benign packages perform behavior patterns that closely resemble malware, such as downloading payloads, decoding them with crypto-related routines, and executing them through \texttt{eval}. In these cases, the LLM may directly report maliciousness without requesting further dynamic validation. We also observe shell-command false positives when installation scripts modify startup configurations or invoke deletion commands against many files, since both patterns resemble common malware persistence and destructive behavior. Separately, false positives arise from source-code paths that invoke downloaded or bundled binaries, which \tool treats as malicious by default because our analysis focuses on source-level behavior rather than validating the semantics of external binaries.}

\tosem{\textbf{Improvement over \textsc{ProfMal}.}}
\tosem{Compared with \textsc{ProfMal}, \tool improves precision from \todo{91.4\%} to \todo{98.0\%}, recall from \todo{90.3\%} to \todo{98.2\%}, and F1-score from \todo{90.9\%} to \todo{98.1\%}, yielding gains of \todo{7.2\%}, \todo{8.7\%}, and \todo{7.9\%}, respectively. The larger improvement in recall suggests that \tool is particularly effective at recovering malicious behaviors missed by \textsc{ProfMal}'s GNN-based classifier. This benefit mainly comes from reasoning over code-level behavior slices, which retain richer semantic evidence than graph-level training features. In \textsc{ProfMal}, the classifier primarily observes node types and sensitivity scores, which may abstract away critical behavioral details such as argument construction, return values, data-flow context, and surrounding control logic. As a result, packages with similar graph structures may still encode substantially different code-level behaviors. By preserving these contextual semantics, \tool can infer malicious intent from surrounding statements, data dependencies, and API usage patterns, rather than relying only on compressed sensitivity signals or behaviors exposed during dynamic execution. The precision gain further indicates that this richer evidence does not simply make \tool more aggressive. Instead, the multi-agent reasoning pipeline helps distinguish malicious behaviors from benign security-relevant operations and reduces over-claiming.}

\tosem{For example, \textsc{ProfMal} falsely reports one benign package whose graph contains repeated platform and release checks, together with a \texttt{child\_process} invocation. Since the concrete command semantics are not resolved in the behavior graph, the command-related node receives a default sensitivity score; combined with the environment-check nodes, the resulting graph becomes close to malicious information-stealing patterns that first fingerprint the host and then execute sensitive operations. The GNN classifier therefore labels it as malicious. However, the corresponding source context shows a common compatibility pattern: the package checks whether it is running on Windows, macOS, Linux, or an unsupported platform, and then loads the matching local implementation. From this source-level view, \tool interprets the behavior as platform-specific dispatch rather than data theft, avoiding this false positive. Conversely, for a \textsc{ProfMal}'s false negative corrected by \tool, one malicious package collects host information such as the hostname, current user, and package path, stores them in HTTP headers, and sends them to an external endpoint through a third-party network module. In this case, the network behavior is implemented through a third-party library. Since the behavior is not exposed during dynamic execution (e.g., the embedded URL has expired and causes an execution error) and the static graph does not recover the semantics of this third-party network call, \textsc{ProfMal}'s graph misses the outbound communication sink and therefore labels the package as benign. Through third-party enrichment, \tool recovers that the third-party call performs an HTTP request, annotates the corresponding source slice with this external API semantics, and reasons jointly over the collected host data, header construction, destination URL, and enriched network sink. This allows \tool to reconstruct the information-exfiltration behavior and correct the false negative.}

\tosem{\textbf{Stage-Wise Decision Flow.}}
\tosem{To understand where \tool reaches its final decisions, Table~\ref{tab:stage_flow} reports the terminal stage of each detection run under \tool’s analysis granularity. For both benign and malicious~packages, \tool analyzes entry files sequentially and stops once any entry is classified as malicious. Therefore, these counts are interpreted as terminal decision flows of detection runs, rather than as package-level totals or counts over all analyzed entry files. Under this accounting, most decisions are made at the static judgement stage; i.e., \todo{89.9\%} of benign flows and \todo{89.5\%} of malicious flows are finalized without invoking later stages. In comparison, \textsc{ProfMal} sends \todo{56.5\%} of all entry files to dynamic analysis. This difference highlights the advantage of using code as the reasoning substrate; i.e., behavior slices preserve concrete API usages, arguments, data dependencies, and local control context, allowing the LLM to directly infer security intent from static evidence instead of waiting for the behavior to be exposed at runtime.}

\begin{table}[!t]
  \centering
  \small
  \caption{\tosem{Stage-Wise Decision Flow of \tool}}
  \vspace{-10pt}
  \label{tab:stage_flow}
    \tosem{%
    \begin{tabular}{
      >{\centering\arraybackslash}m{2.5cm}
      >{\centering\arraybackslash}m{2.5cm}
      >{\centering\arraybackslash}m{2.5cm}
      >{\centering\arraybackslash}m{2.5cm}
      >{\centering\arraybackslash}m{2.5cm}
    }
      \toprule
      \textbf{Dataset} & \textbf{Static Only} & \textbf{Static + Dyn. Aug.} & \textbf{Static + Third-Party Enrich.} & \textbf{Static + Third-Party Enrich. + Dyn. Aug.} \\
      \midrule
      Benign    & 3,698 (89.9\%) & 250 (6.1\%) & 120 (2.9\%) & 44 (1.1\%) \\
      Malicious & 1,015 (89.5\%) & 96 (8.5\%)  & 23 (2.0\%)  & 0 (0.0\%)  \\
      \bottomrule
    \end{tabular}%
    }
\end{table}

\tosem{\textbf{Time Overhead.}}
\tosem{For runtime efficiency, \textsc{Malpacdetector} is the fastest baseline, requiring only \todo{1.44}~s/pkg, followed by \textsc{Cerebro} at \todo{25.64}~s/pkg. \textsc{GuardDog} takes \todo{151.88}~s/pkg, while \textsc{SocketAI} and \textsc{ProfMal} take \todo{176.91}~s/pkg and \todo{192.71}~s/pkg, respectively. \tool takes \todo{394.37}~s/pkg, which is higher than all baselines. This overhead mainly comes from LLM invocations in the multi-agent pipeline. Unlike traditional classifiers that make a single prediction over extracted features, \tool spends additional time collecting and reasoning over richer semantic evidence. This design trades runtime efficiency for higher detection effectiveness.}

\tosem{\textbf{Monetary Cost.}}
\tosem{Regarding monetary cost, \textsc{SocketAI} costs \todo{\$562.48} on the evaluation dataset, while \tool costs only \todo{\$67.45}, making \tool \todo{8.3} times cheaper while also improving F1-score by \todo{3.5\%}. The main reason is that \textsc{SocketAI} sends whole package files to the LLM, which can be viewed as a degenerated version of LLM-based code analysis without behavior-guided reduction. In contrast, \tool first extracts behavior-relevant code slices and invokes the LLM only on the semantic evidence needed for judgement, avoiding exhaustive analysis of irrelevant package code. Compared with \textsc{ProfMal}, \tool costs more (\todo{\$67.45 vs. \$19.43}) because the extended pipeline invokes LLM reasoning in more stages, but the additional cost is substantially smaller than whole-file LLM analysis.}

\begin{tcolorbox}[size=title, opacityfill=0.15]
  \tosem{\textbf{Summary.} \tool achieves the highest precision, recall, and F1-score among all evaluated detectors. Its main gain over \textsc{ProfMal} comes from replacing the GNN classifier with evidence-driven multi-agent reasoning, which improves both recall and precision. This richer reasoning incurs higher time overhead due to additional LLM invocations, while remaining substantially cheaper than \textsc{SocketAI} in monetary cost.}
\end{tcolorbox}

\subsection{\tosem{LLM Backbone Selection (RQ2)}}

\tosem{\textbf{RQ2 Setup.} \tool's multi-agent pipeline invokes an LLM across multiple stages, including shell-command analysis, third-party enrichment, local and global behavior judgement, routing, and localization. We compare DeepSeek-V4-Flash~\cite{deepseek-v4-Flash} with GPT-5.4-mini~\cite{gpt-5-4-mini} and Qwen-3.6-Flash~\cite{qwen-3-6-flash} as representative lightweight models from major commercial and open-source providers. To compare their effectiveness and cost tradeoff, we fix the entire \tool pipeline, and only replace the LLM backend. For each backbone, we report the precision, recall, and F1-score, together with the total cost. All other hyperparameters, prompts, and agent orchestration logic are kept identical across backbones.}

\begin{table}[!t]
  \centering
  \small
  \caption{\tosem{Results of LLM Backbone Selection}}
  \vspace{-10pt}
  \label{tab:llm_backbone}
  \tosem{%
    \begin{tabular}{
      >{\centering\arraybackslash}m{2.6cm}
      >{\centering\arraybackslash}m{1.5cm}
      >{\centering\arraybackslash}m{1.5cm}
      >{\centering\arraybackslash}m{1.5cm}
      >{\centering\arraybackslash}m{1.5cm}
    }
    \toprule
    \textbf{LLM Backbone} & \textbf{Precision} & \textbf{Recall} & \textbf{F1-Score} & \textbf{Total Cost} \\
    \midrule
    DeepSeek-V4-Flash  & 98.0\% & 98.2\% & 98.1\% & \$67.45 \\
    GPT-5.4-mini       & 93.8\% & 93.2\% & 93.6\% & \$235.81 \\
    Qwen-3.6-Flash     & 98.1\% & 94.6\% & 96.3\% & \$314.80 \\
    \bottomrule
    \end{tabular}
    }
\end{table}

\tosem{\textbf{Overall LLM Comparison Results.} Table~\ref{tab:llm_backbone} summarizes the effectiveness and cost tradeoff across the three~lightweight backbones. DeepSeek-V4-Flash achieves the best overall F1-score (\todo{98.1\%}) and recall (\todo{98.2\%}) while incurring the lowest monetary cost (\todo{\$67.45}), making it the most favorable default backbone. Qwen-3.6-Flash obtains a slightly higher precision (\todo{98.1\%}), but its lower recall (\todo{94.6\%}) reduces the F1-score to \todo{96.3\%}, and its total cost rises to \todo{\$314.80}. GPT-5.4-mini performs worse on all three effectiveness metrics, with \todo{93.8\%} precision, \todo{93.2\%} recall, and \todo{93.6\%} F1-score, while costing \todo{\$235.81}. Therefore, higher-cost backbones do not provide a corresponding effectiveness gain in this setting, and DeepSeek-V4-Flash is used as the default backbone in RQ1 and RQ3--RQ5.}

\begin{tcolorbox}[size=title, opacityfill=0.15]
\tosem{\textbf{Summary.} Under the same \tool pipeline, DeepSeek-V4-Flash achieves the highest F1-score and recall at the lowest monetary cost, whereas the more expensive GPT-5.4-mini and Qwen-3.6-Flash do not improve the overall detection tradeoff.}
\end{tcolorbox}

\subsection{Ablation Study (RQ3)}

\tosem{\textbf{RQ3 Setup.}}
\tosem{We evaluate the contribution of each key component in \tool's pipeline by constructing four ablated versions. (1) \emph{w/o MSC-Detection} removes only the malicious shell command detection from the script analyzer (Sec.~\ref{sec:scripts_analyzer}), while keeping the entry file extraction and third-party dependency parsing intact, so that the subsequent behavior-graph construction and reasoning pipeline are not affected. (2) \emph{w/o Third-Party Enrich.} removes the \texttt{third\_party\_enrichment} action from the router agent, so that uncertainty caused by third-party calls can only fall through to dynamic augmentation (Sec.~\ref{sec:third_party_enrichment}). (3) \emph{w/o Dyn. Aug.} removes the \texttt{dynamic\_augmentation} action, so that the runtime evidence behind routed \emph{conditional}, \emph{eval}, and statically unknown nodes can no longer be collected; consequently, any entry that stays \texttt{undetermined} and would require dynamic evidence to reach a verdict is conservatively classified as benign (Sec.~\ref{sec:behavior_graph_dynamic}). (4) \emph{w/o Self-Consistency} disables self-consistency verification within the local behavior judge agent, invoking the agent once per slice rather than sampling three candidate reports and consolidating them with the verifier (Sec.~\ref{subsub:local_judge}). All other components and the LLM backbone are kept identical across versions.}

\begin{table}[!t]
  \centering
  \small
  \caption{\tosem{Results of Ablation Study}}
  \vspace{-10pt}
  \label{tab:ablation_study}
    \tosem{%
    \begin{tabular}{ 
      >{\centering\arraybackslash}m{4.5cm} 
      >{\centering\arraybackslash}m{1.4cm} 
      >{\centering\arraybackslash}m{1.4cm}
      >{\centering\arraybackslash}m{2.0cm}
      }
    \toprule
    \textbf{Ablated Version} & \textbf{Precision} & \textbf{Recall} & \textbf{F1-Score} \\
    \midrule
    \tool                             &   98.0\%   &   98.2\% &  98.1\%               \\
    \tool w/o MSC-Detection           &   97.8\%   &   84.0\% &  90.4\% ($\downarrow$ 7.85\%) \\
    \tool w/o Third-Party Enrich.     &   98.1\%   &   97.4\% &  97.8\% ($\downarrow$ 0.30\%) \\
    \tool w/o Dyn. Aug.               &   98.1\%   &   90.7\% &  94.3\% ($\downarrow$ 3.87\%) \\
    \tool w/o Self-Consistency        &   89.5\%   &   98.4\% &  93.7\% ($\downarrow$ 4.49\%) \\
    \bottomrule
    \end{tabular}%
    }
  \end{table}

\tosem{\textbf{Overall Ablation Results.}}
\tosem{Table~\ref{tab:ablation_study} presents the results of our ablation study. Eliminating any individual component degrades the overall effectiveness, confirming that each ablated component contributes to \tool. Eliminating the malicious shell command detection results in a \todo{7.85\%} decrease in F1-score, primarily due to reduced recall, because several malicious packages in the dataset inject malicious shell commands during the installation phase that are no longer caught by the script analyzer.}

\tosem{Eliminating the dynamic augmentor causes a \todo{3.87\%} drop in F1-score, dominated by a recall decrease from \todo{98.2\%} to \todo{90.7\%}. The overall impact is modest because, as shown in Table~\ref{tab:stage_flow}, nearly \todo{90\%} of detection runs already reach their verdicts at the static judgement stage. The degradation instead concentrates on the packages whose maliciousness can only be confirmed at runtime, such as payloads hidden behind \texttt{eval}, or runtime-dependent conditions. Without dynamic evidence, the corresponding entries stay \texttt{undetermined} and are conservatively classified as benign, so these malicious packages are missed.}

\tosem{Disabling self-consistency verification within local judging causes a \todo{4.49\%} drop in F1-score, almost entirely driven by precision, which falls from \todo{98.0\%} to \todo{89.5\%} while recall stays essentially unchanged (\todo{98.2\%} to \todo{98.4\%}). Without this verification, each local decision relies on a single stochastic LLM judgement. If that single sample over-interprets incomplete evidence as malicious, there is no follow-up check to downgrade the result to \texttt{undetermined}. The verifier reduces these one-shot errors by consolidating multiple independently sampled judgements and checking their evidence~\cite{wang2023selfconsistency}. Once self-consistency verification is disabled, many benign packages with weak or ambiguous evidence are incorrectly labeled as malicious, increasing the number of false positives.}

\tosem{Removing the third-party enrichment agent leads to only a \todo{0.3\%} drop in F1-score, which is smaller than the impact of other ablated components. This is because the router can still fall back to dynamic augmentation and recover most third-party semantics at runtime. The degradation is mainly reflected in recall. After enrichment is disabled, third-party uncertainties are routed to dynamic augmentation, which reproduces most relevant behaviors during execution. As a result, only a small number of malicious packages whose third-party behaviors are not triggered dynamically become false negatives. Therefore, the main contribution of third-party enrichment is not improving raw effectiveness, but improving efficiency and selective evidence acquisition. In the complete \tool pipeline, the router invokes enrichment on demand for \todo{187} detection runs. Among them, \todo{143} cases (\todo{76.5\%}) are resolved using documentary evidence alone and do not require subsequent dynamic augmentation, while the remaining \todo{44} cases (\todo{23.5\%}) fall through to dynamic augmentation because the recovered semantics are untrusted, too generic, or insufficient. This result shows that enrichment serves as an effective but non-terminal evidence source.}

\tosem{We further quantify this efficiency benefit by comparing the average runtime of enrichment and dynamic augmentation under the same evaluation setting. Resolving a third-party uncertainty through enrichment takes \todo{61.76}s on average, because it only retrieves registry information, documentation, and declaration files, followed by a few lightweight LLM calls. In contrast, one dynamic execution in the instrumented sandbox takes \todo{357.43}s on average, making it about \todo{5.8} times slower. Thus, cases resolved by enrichment alone require only about \todo{17.3\%} of the runtime of sandbox execution. Enrichment may introduce extra cost for fall-through cases; i.e., when the recovered semantics are still insufficient and dynamic augmentation is needed, the sandbox is invoked after enrichment, increasing the average runtime to \todo{406.58}s, about \todo{49}s more than direct dynamic execution. However, such fall-through cases are the minority, and the extra cost is small compared with a full sandbox run. Overall, these results show that \tool benefits from first resolving third-party semantics with cheap documentary evidence and invoking dynamic augmentation only when necessary.}

\begin{tcolorbox}[size=title, opacityfill=0.15]
  \tosem{\textbf{Summary.} Each ablated component contributes to \tool's overall effectiveness. Since most verdicts are already reached at the static judgement stage, removing the dynamic augmentor causes only a modest degradation, but it is indispensable for confirming packages whose malicious behaviors are only observable at runtime. The third-party enrichment agent mainly improves efficiency and selectivity by resolving third-party semantics with cheap documentary evidence and reducing reliance on costly dynamic execution.}
\end{tcolorbox}

\subsection{\tosem{Interpretability Evaluation (RQ4)}} \label{sec:interpretability-evaluation}

\tosem{\textbf{RQ4 Setup.}
We evaluate interpretability on confirmed true-positive malicious packages. Specifically, from the \todo{1,070} true positives reported by \tool, we randomly sample \todo{283} packages for manual evaluation, corresponding to a \todo{95\%} confidence level and a \todo{0.05} margin of error under finite population correction.}

\tosem{Two security experts independently inspect each sampled package and \tool's output from two perspectives. First, they annotate the minimal source code lines that implement the malicious behavior. The annotation unit~is~a~file-line pair, so disjoint malicious fragments in the same package can be labeled separately. Experts are instructed to include lines necessary to understand the malicious behavior, such as payload construction, sensitive source access,~encoding or decoding, guard conditions, and the final sensitive sink, but to exclude unrelated helper code and benign surrounding context. Second, they assess the quality of the malicious-behavior explanation produced by \tool. Each~expert identifies the malicious behavior type based on manual code inspection, checks whether this type is reflected~in~\tool's generated summary, and rates the quality of the explanation. After independent annotation and assessment, disagreements are discussed and resolved into a consensus oracle. To quantify the reliability of this manual oracle before reconciliation, we report inter-expert agreement for source line annotation and explanation quality rating.}

\tosem{\textbf{Line-Level Localization Metric.}
Before reconciliation, we compute inter-expert line-level F1-score by comparing the two experts' file-line annotation sets for each package and averaging the resulting F1-scores across sampled~packages. Specifically, given the two experts' annotation sets $L_a$ and $L_b$, we compute $2 \times |L_a \cap L_b|/(|L_a|+|L_b|)$ for each package. For \tool's localization result, we convert its reported localization snippets into a set of file-line pairs $L_t$, and compare them with the expert consensus set $L_g$. We measure line-level precision, recall, and F1-score by
$P_{line}=|L_t \cap L_g|/|L_t|$, $R_{line}=|L_t \cap L_g|/|L_g|$, and $F1_{line}=2 \times P_{line} \times R_{line}/(P_{line}+R_{line})$. We compute these metrics per package and report the average across the sampled packages.}

\tosem{\textbf{Malicious Behavior Explanation Metric.}
We evaluate whether \tool's explanation is useful for human auditing by scoring its overall explanation quality. For each sampled package, the experts first identify the malicious behavior type through manual code inspection. For example, if the experts consider a package to perform credential collection, they judge whether this behavior type is reflected in \tool's generated summary. The experts~also assess whether the summary explains why the behavior is malicious, provides behavior-specific details, and is supported by the localized code snippets.}
\tosem{Specifically, the experts rate explanation quality on a three-level scale. A score of 0 means the explanation is incorrect, benign-oriented, or unsupported by the localized code. A score of 1 means the explanation captures the malicious intent or broad behavior type, but misses important details, such as the sensitive source, destination, trigger condition, payload, or security effect, or contains partially unsupported claims. A score of 2 means the explanation correctly identifies the malicious behavior type, explains why the behavior is malicious with sufficient behavior-specific details, and each major claim is supported by the localized snippet. For example, an exfiltration explanation should identify what data is collected and where it is sent, while a download-and-execute explanation should identify the payload source and the execution sink. Before reconciliation, we report Cohen's $\kappa$ for the two experts' explanation quality ratings. After reconciliation, we report the average explanation quality score as well as the high-quality explanation rate (i.e., the fraction of packages whose explanation receives a score of 2).}

\begin{table}[!t]
  \centering
  \small
  \caption{\tosem{Results of Interpretability Evaluation}}
  \vspace{-10pt}
  \label{tab:interpretability}
    \tosem{
    \begin{tabular}{
      >{\centering\arraybackslash}m{4.4cm}
      >{\centering\arraybackslash}m{1.8cm}
      }
    \toprule
    \textbf{Metric} & \textbf{Result} \\
    \midrule
    Inter-Expert Line-Level F1                & 92.4\%            \\
    Inter-Expert Explanation $\kappa$   & 0.79              \\
    Line-Level Localization F1          & 88.9\%            \\
    Explanation Quality Score (0--2)    & 1.87              \\
    High-Quality Explanation Rate       & 86.9\%            \\
    \bottomrule
    \end{tabular}
    }
\end{table}

\tosem{\textbf{Overall Interpretability Results.}
Table~\ref{tab:interpretability} summarizes the interpretability evaluation. The two experts achieve a \todo{92.4\%} line-level F1 and a Cohen's $\kappa$ of \todo{0.79} for explanation quality ratings before reconciliation, indicating reliable manual annotations. Compared with the reconciled oracle, \tool achieves an \todo{88.9\%} line-level localization F1-score, showing that its reported snippets usually cover the malicious source lines selected by human experts. For explanation quality, \tool obtains an average score of \todo{1.87} on the 0--2 scale, and \todo{86.9\%} of the sampled packages receive a high-quality explanation score of 2. The remaining partially useful explanations typically identify the broad malicious behavior but omit a key source, sink, trigger, or security effect.}

\tosem{\textbf{Localization Deviation Analysis.}
We further inspect line-level over-coverage and under-coverage. Over-coverage mainly comes from two sources. First, some extracted code patterns are general JavaScript idioms rather than malware-specific syntax. For example, benign helper routines may use the same decoding or string-construction idioms as malicious snippets, such as base64 decoding through \texttt{Buffer.from(...).toString()}, and can therefore be included when matching localized code fragments. Second, context expansion around suspicious call nodes can exceed the true injection boundary. When malicious code is placed next to benign utility code, the expanded snippet may include surrounding legitimate synchronization, filesystem, or I/O statements. Under-coverage mainly arises because the localization agent tends to preserve the core malicious actions while omitting peripheral but still security-relevant lines. Trigger conditions, such as platform checks, are often part of the malicious execution path, but they can be ignored when the localized snippet starts from the subsequent malicious operation. Similarly, tail code for cleanup or persistence, such as removing temporary files or writing auxiliary state, may be omitted when it appears less central than the main malicious action. Finally, import statements for sensitive modules, such as \texttt{child\_process}, \texttt{net}, and \texttt{crypto}, are frequently omitted because they prepare the attack but do not execute the malicious action themselves, even though experts include them as infrastructure lines for the malicious behavior.}

\tosem{\textbf{Explanation Deviation Analysis.}
For partially useful explanations, \tool generally identifies the correct malicious behavior type, but sometimes lacks behavior-specific details. In data-theft cases, the summary may state that sensitive information is exfiltrated without clearly identifying the collected source data or the outbound sink. In download-and-execute cases, it may capture the high-level behavior but omit the payload source, execution target, or trigger condition. In addition, some explanation gaps are caused by incomplete code extraction; i.e., when the localized snippet misses supporting lines, the generated summary has insufficient evidence to describe the full attack chain.}

\begin{tcolorbox}[size=title, opacityfill=0.15]
  \tosem{\textbf{Summary.} \tool achieves accurate malicious code localization and produces high-quality explanations for most sampled malicious packages. These results demonstrate that \tool effectively supports human auditors in locating malicious logic and understanding its security impact.}
\end{tcolorbox}

\subsection{Usefulness Evaluation (RQ5)} \label{sec:usefulness-evaluation}

\textbf{RQ5 Setup.}
We deployed all the detectors except~for~\textsc{SocketAI} in a real-world setting to detect newly published NPM packages. \textsc{SocketAI} incurred a cost of \todo{\$562.48} in our effectiveness~evaluation, making it impractical for real-world detection due to the high expense. \tosem{For \textsc{EMPHunter}, we followed its original real-world detection setting~\cite{liang2025ClusteringInstallationScripts}; i.e., newly published packages were aggregated into batches, their installation scripts were deduplicated by canonical sequence, each batch was clustered with benign NPM installation scripts, and the top-10 ranked outliers in each batch were treated as suspicious candidates for manual review.} We monitored all newly published NPM packages every five minutes between \tosem{April 2026 and June 2026}, collecting a total of \todo{107,287} packages for real-world detection.

\begin{table}[!t]
  \centering
  \small
  \caption{Results of Usefulness Evaluation}
  \vspace{-10pt}
  \label{tab:real-world}
    \tosem{
    \begin{tabular}{c
      >{\centering\arraybackslash}m{1.0cm}
      >{\centering\arraybackslash}m{1.3cm}
      >{\centering\arraybackslash}m{1.3cm}
      >{\centering\arraybackslash}m{1.3cm}
      >{\centering\arraybackslash}m{1.3cm}
      >{\centering\arraybackslash}m{1.3cm}}
      \toprule
      \textbf{Tool} & \textbf{Detected} & \textbf{TP} & \textbf{FP Rate} & \textbf{FN} & \textbf{FN Rate} \\
      \midrule
       \textsc{GuardDog}       &  4,687 &   567  & 87.9\%  &  91  & 13.8\%  \\
       \textsc{Cerebro}        &  2,034 &   505  & 75.2\%  &  153 & 23.3\%  \\
       \textsc{EMPHunter}      &  663   &   146  & 78.0\%  &  512 & 77.8\%  \\
       \textsc{Malpacdetector} &  1,621 &   423  & 73.9\%  &  235 & 35.7\%  \\
       \textsc{ProfMal}        &  736   &   527  & 28.4\%  &  131 & 19.9\%  \\
       \tool                   &  715   & \textbf{597}   & \textbf{16.5\%}  & \textbf{61} & \textbf{9.3\%}  \\
      \bottomrule              
    \end{tabular}
    }
\end{table}

\textbf{Overall Usefulness Results.} \tosem{Table~\ref{tab:real-world} presents the real-world detection results. \tool detected \todo{715} potentially~malicious packages, of which \todo{597} were confirmed as malicious after~manual review. These packages have been confirmed as malicious by NPM and subsequently removed. Among all detectors, \tool identified the highest number of malicious packages while achieving the lowest false positive rate of \todo{16.5\%}. Compared to state-of-the-art detectors, \tool reduced the false positive~rate~by~\todo{41.9\%~to~81.2\%}. Besides, as NPM did not disclose a complete list of malicious packages, we used all the malicious packages detected by all detectors to compute false negatives. \tool also achieved the lowest false negative rate of \todo{9.3\%}. \textsc{GuardDog} had the lowest false negative rate among all baselines, but its false positive rate was extremely high. These results demonstrate the practical usefulness of \tool in the real world.}
 
\tosem{\textbf{False Positive Analysis.}
We further summarize the main causes of false positives. First, most false positives are related to binary execution. \tool reasons about package behaviors from source-code-level API usage. When a package executes a local binary or downloads an executable from a remote server, the actual behavior is hidden inside an opaque binary and cannot be inspected by \tool. Therefore, \tool conservatively treats such behaviors as high-risk. This design is safety-oriented, but it can also misclassify benign binary execution or benign executable downloads as malicious. A possible direction is to combine \tool with system-call-level analysis, which can observe the runtime behavior of binaries and provide more precise evidence. Second, some packages establish remote interactive servers or execute remote commands with elevated privileges. Although their dynamic execution fails in our environment, these behaviors remain highly sensitive and are therefore flagged as malicious. Third, some legitimate packages access sensitive user data as part of their intended functionality, but are reported as malicious because their behavior matches a sensitive source-to-sink pattern. For example, a large language model deployment tool reads the user’s token, which forms a token-upload behavior, but this access is benign in context. Similar cases include Facebook account management tools and wallet management tools, where reading credentials or wallet information is expected. Future work should incorporate repository-wide information, such as the package’s declared purpose and surrounding code context, to further reduce this class of false positives.}

\tosem{\textbf{False Negative Analysis.}
We also summarize the main causes of false negatives. First, \tool focuses on behaviors triggered during installation and import time. Therefore, malicious code that is activated only through explicit user invocation at runtime falls outside the current analysis scope, leading to false negatives. Second, some false negatives are an inherent cost of \tool’s conservative decision policy, which favors a low false positive rate over aggressive reporting. \tool assigns the malicious verdict only when sufficient node-grounded evidence is available. As a result, if neither third-party enrichment nor dynamic execution can expose decisive dynamic evidence, the reasoning loop terminates with a non-malicious verdict rather than reporting a package based on weak evidence. This situation occurs, for example, when an obfuscated payload is activated only on Windows or guarded by specific conditional checks, such as checking for particular sensitive-data keys. In such cases, our Linux-based dynamic analysis may not observe the malicious behavior. Similar false negatives can also occur when the decisive payload, URL, or dependency is unavailable at execution time.}

\tosem{This real-world setting further offers a temporal separation for evaluating an LLM-based detector. Because all packages analyzed in RQ5 were newly published during our monitoring period, they postdate the training cutoff of the LLM backbone and thus are unlikely to have been seen in its training corpus. Thus, \tool largely had to reach its verdicts from the observed code and behavioral evidence rather than from recall of previously known malicious packages. This suggests that its effectiveness is driven primarily by genuine reasoning rather than by~data~leakage.}

\begin{tcolorbox}[size=title, opacityfill=0.15]
  \tosem{\textbf{Summary.} \tool identified \todo{597} new malicious packages in NPM over three months, achieving the lowest false positive rate and false negative rate among state-of-the-art detectors, demonstrating its practical usefulness in real-world detection.}
\end{tcolorbox}


\section{Discussion} 

\subsection{Threats to Validity}
Although dynamic analysis can augment static~analysis, not all packages in our dataset were executable. Typical reasons include missing third-party commands in the standard Ubuntu environment, invalid URLs (often taken down after the disclosure of malicious packages), unavailable dependencies (no longer maintained or downloadable), platform constraints, and missing environment variables or user files (e.g., wallets). We only excluded obfuscated packages that failed to execute to ensure runnable experiments, since analyzing these excluded packages would have almost entirely relied on dynamic evidence under our evaluation setting. These removed packages might exhibit different malicious behaviors compared to the retained set. Besides, we manually analyzed malicious packages to filter out those whose malicious behaviors were triggered at run-time. This manual step introduced subjectivity, and the filtering may reduce diversity. \tosem{Moreover, \tool builds on LLMs whose outputs are inherently non-deterministic, so the same slice may receive different verdicts across invocations. We mitigate this variance with self-consistency verification, but cannot fully eliminate it. The line-level localization oracle and the malicious/benign labels used in the real-world study were also established through manual inspection, which is unavoidably subjective. To reduce this threat, two experts annotated each sample independently and reconciled disagreements into a consensus, and we report inter-expert line-level F1 and Cohen's $\kappa$ to quantify their agreement before reconciliation. Finally, since NPM does not publish a complete ground-truth list of malicious packages, the real-world false negatives in RQ5 are computed against the union of all detectors' confirmed detections rather than the true population of malicious packages, which may underestimate the absolute false negatives while still enabling a consistent relative comparison.}

\subsection{Limitations}
No detector can be considered a complete and isolated solution~\cite{ohm2023sok}. Likewise, there are some limitations of \tool. First, \tool currently targets behaviors~executed during the installation and import time. Although most malicious~behaviors occur within these phases, attacks activated only at run-time fall outside this scope and evade detection. Second, the fidelity of our static modeling is bounded: despite employing object-sensitive analysis, \tool does not address prototype pollution~\cite{shcherbakov2023silent}, enabling attackers to bypass detection through prototype manipulation, \modified{and it relies on third-party tools whose inaccuracies may propagate into our analysis.} Third, some behaviors remain opaque to source-level reasoning. \tool conservatively flags executable~files as malicious in both code behavior and shell~command analysis, and its dynamic analysis depends on successful execution under Linux, which some packages may not satisfy; both cases may lead to false positives or false negatives and motivate complementary system-level analysis in future. \tosem{Fourth, the LLM reasoning stage has its own constraints. Our RQ2 evaluation compares only mainstream lightweight backbones to balance detection quality with deployment cost, and does not assess heavier frontier models that might better reason over complex multi-stage payloads or subtle contextual dependencies; so the reported effectiveness may not represent the upper bound attainable with more powerful LLMs. Moreover, although code slicing substantially reduces the code fed to the LLM by retaining only behavior-relevant statements, very large files can still produce slices that approach or exceed the context window, in which case the truncated evidence may weaken the judgement and cause missed behaviors.}

\subsection{Future Directions}

\tosem{We outline several directions that we think are promising for the next generation of supply-chain malware detectors.} 

\tosem{\emph{(1) Runtime-aware lifecycle coverage.} Recent studies have started to explore runtime behaviors through techniques such as fuzzing or synthesized inputs. This trend suggests that future detectors should no longer treat installation and import phases as the only primary observation points. Instead, runtime execution should become a first-class part of supply-chain malware detection, because malicious logic may be delayed until explicit user invocation, specific API calls, or particular environmental conditions are satisfied. Building detectors that systematically cover installation, import, and runtime phases would help reduce this blind spot and provide a more complete view of package behavior.}

\tosem{\emph{(2) Holistic execution monitoring.} Many packages obtain and execute external artifacts, such as locally bundled binaries or executables downloaded from remote servers. These artifacts may hide behaviors that are difficult to infer from~source code alone. A promising direction is to build holistic monitoring across the full execution process, covering not only high-level API calls but also deeper behaviors such as process creation, file-system effects, network communication, privilege changes, and system calls. Such execution-stage monitoring would allow future tools to reason about both visible source-level behaviors and hidden effects introduced by external artifacts, leading to more comprehensive behavioral evidence.}

\tosem{\emph{(3) Context-aware judgement.} Some false positives stem from benign packages whose intended functionality legitimately touches sensitive data, such as credential managers or wallet management tools. Incorporating repository-wide context, including the declared purpose, documentation, package metadata, and surrounding code, would help distinguish expected sensitive behavior from genuine abuse.}

\tosem{\emph{(4) AI-native malware detection.} Future detectors can further exploit the reasoning, planning, and tool-using capabilities of large language models. Rather than using LLMs only as isolated classifiers, an AI-native detector could organize multiple specialized agents for complementary tasks, such as code understanding, behavior reconstruction, dynamic evidence interpretation, package-intent inference, and final risk assessment. These agents can collaborate with static analyzers, dynamic sandboxes, binary monitors, and external knowledge sources to form an adaptive reasoning pipeline. Such a design would allow the detector to spend more effort on genuinely ambiguous cases while maintaining efficiency on simple cases, providing a path toward stronger and more explainable supply-chain malware detection.}


\section{Related Work}


Several empirical studies \cite{ohm2020backstabber, zhou2024largeScaleAnalysis, guo2023empiricalOfPyPI, Ladisa2023HitchhikerGuide, zhou2024MaliciousPackagesInWild, zhang2024TTPs} have~been conducted to understand malicious packages in various ecosystems. These studies provide good insights into designing~detectors. Various approaches have also been developed to detect malicious NPM packages, and these tools can be broadly categorized into four types, rule-based~\cite{huang2024spiderscan,li2023malwukong,zheng2024OSCAR,Pohl2024runtimeProtect,zahan2022weakLinks,gonzalez2021maliciousCommits,taylor2020defendingTypo,wyss2022wolfAtDoor,duan2020maloss,GuardDog,OSSBackdoor}, learning-based~\cite{yu2024maltracker,nguyen2024classifyingByDynamic,Halder2024metainfo,huang2024donapi,zhang2024cerebro,ladisa2023crossLanguage,sejfia2022amalfi,ohm2022supervisied,ohm2022malevolent,garrett2019detectingUpdate,liang2025ClusteringInstallationScripts,wang2025Malpacdetector},  LLM-based~\cite{zahan2024detectionByLLM}, and differential testing-based~\cite{sofaer2024rogueone,scalco2022injectionInNpm}.

Rule-based approaches rely on predefined rules to analyze various package aspects, e.g., metadata \cite{zahan2022weakLinks, taylor2020defendingTypo, duan2020maloss}, static~behaviors \cite{li2023malwukong, GuardDog, OSSBackdoor, duan2020maloss}, and runtime behaviors \cite{duan2020maloss}. 
\textsc{MalOSS}~\cite{duan2020maloss} combines metadata, static behaviors, and dynamic behaviors.
Several protection tools~\cite{Pohl2024runtimeProtect, wyss2022wolfAtDoor} defend malicious behaviors at runtime, but are not originally designed for malicious package detection. While rule-based approaches are simple, they suffer high false positives. To mitigate this, \textsc{SpiderScan}~\cite{huang2024spiderscan} refines sensitive APIs into fine-grained types, builds behavior graphs, and performs dynamic analysis on targeted APIs to confirm maliciousness; and \textsc{OSCAR}~\cite{zheng2024OSCAR} generates API~behavior sequences dynamically, and applies enhanced system-level~rules.

Learning-based approaches leverage metadata \cite{Halder2024metainfo, sejfia2022amalfi}, static behaviors \cite{yu2024maltracker, sejfia2022amalfi, ladisa2023crossLanguage}, and dynamic behaviors \cite{nguyen2024classifyingByDynamic} as features to learn a maliciousness classifier. Differently, Zhang et al.~\cite{zhang2024cerebro} and Huang et al.~\cite{huang2024donapi} propose to use API sequences to learn the behavior. Liang et al.~\cite{liang2025ClusteringInstallationScripts} extract function call sequences from the control-flow graph and normalize the traversal so that semantically similar code yields similar token sequences, then cluster benign installation scripts and flag outliers as malicious. Wang et al.~\cite{wang2025Malpacdetector} use an LLM to summarize malicious behavior features from known malicious packages and then train a conventional classifier on these features, locating evidence via feature matching and AST node mapping. Ohm et al.~\cite{ohm2022supervisied} evaluate the performance of various machine learning models in malicious package detection.  
Recently, Zahan et al.~\cite{zahan2024detectionByLLM} introduced \textsc{SocketAI} to investigate the potential of LLMs for malicious code detection. Building on this line of work, knowledge-driven retrieval-augmented approaches have been explored to incorporate external threat intelligence into LLM-based detection pipelines~\cite{guo2026bridging}, enabling context-aware reasoning over malicious code behaviors.

These approaches suffer from imprecise modeling of program behavior as summarized in Sec.~\ref{sec:intro} and motivated in Sec.~\ref{sec:motivations}. It is also worth mentioning that differential testing-based approaches \cite{sofaer2024rogueone, scalco2022injectionInNpm} identify discrepancies between source~code and distributed artifacts or between updates.~In~contrast,~\tool only analyzes the source code of individual packages.

\section{Conclusions}

\tosem{We have presented \tool, a malicious NPM package detector that reasons over source-level code slices with a coordinated multi-agent framework. From a unified behavior graph built by the synergy of static and dynamic analysis, \tool extracts behavior-relevant code slices as the analysis target, preserving the original source semantics rather than abstracting them into discrete features. It then coordinates multiple agents that reason over local and global behaviors and adaptively decide when to gather additional evidence, yielding an interpretable verdict with concrete malicious-code localization. Our experiments demonstrate its effectiveness and practical usefulness, achieving the highest F1-score among five state-of-the-art detectors and uncovering \todo{597} previously unknown malicious packages during real-world monitoring.}

\section*{Acknowledgment}
This work was supported by the National Natural Science Foundation of China (Grant No. 62332005 and 62372114).

\bibliographystyle{ACM-Reference-Format}
\bibliography{src/references}

\end{document}